\documentclass[aps,prb,twocolumn,longbibliography]{revtex4-2}
\usepackage{graphicx}
\usepackage{hyperref}
\usepackage{xcolor}
\usepackage{amsmath}
\usepackage{amsfonts}
\usepackage{amssymb}
\usepackage{varioref}
\usepackage{float}
\usepackage{dcolumn}   
\usepackage{subfigure}
\usepackage{epstopdf}
\usepackage{braket}
\usepackage{physics}
\usepackage{hyperref}
\hypersetup{colorlinks=true, 
	urlcolor=blue,
	citecolor=blue, 
	linkcolor=blue}

\begin{document}
	\textheight=23.8cm

\title{Role of Noise on Defect Formation and Correlations in a Long-Range Ising Model Under Adiabatic Driving}
\author{ Santanu Dhara and Suhas Gangadharaiah}
\affiliation {Department of Physics, Indian Institute of Science Education and Research, Bhopal, India}

\date{\today}

\begin{abstract}
 We study an exactly solvable long-range (LR) transverse-field Ising model (TFIM) with a power-law decaying interaction characterized by a decay exponent $\alpha$. In the thermodynamic limit, the system is adiabatically driven in the presence of noise, from paramagnetic phase with all spins down to one with all spins up. Our study examines the role of long-range interactions on the defect density, its distribution, and spin correlations, comparing noisy and noiseless scenarios. In the noiseless case, within the long-range regime, the steady-state properties are primarily influenced by modes near the $k=\pi$ region. However, in the presence of noise, the dominant
contributions shift to the modes near $k=0$. This differs from the SR model, where previous
studies have shown that modes around $k=\pi/2$ play a significant role under noisy condition.
In the absence of noise, defect density scales as $n\propto \tau_Q^{-1/2}$, implying scaling exponent independent of decay exponent. However, we find that decreasing the value of $\alpha$ (i.e., increasing the range) enhances the defect density, whereas in the presence of noise, it is suppressed. In the LR regime, two-point fermionic correlators initially exhibit Gaussian decay, followed by quadratic suppression instead of power-law decay for both noisy and noiseless scenarios. Meanwhile, spin correlators, expressed as a string of fermionic operators, undergo purely exponential decay with no crossover behavior. Furthermore, our analysis of defect formation reveals the influence of LR interaction on the kink-number distribution and its cumulants.
\end{abstract}

\maketitle

\section{Introduction}
 Recent studies on quantum long-range (LR) systems has garnered significant attention due to the need to understand the fundamental physics of nonlocal interactions and their effects on key properties like entanglement, coherence, and correlation length~\cite{Campa2009, NicoloDefenu2023}. Long ranged systems also play a key role as powerful tools for efficient quantum computing and simulations~\cite{Blatt2012,Votto2024universalquantum,Browaeys2020, Yao2021}. Recent experimental advancements, particularly in ultracold atomic systems, have enabled unprecedented control over quantum systems, with platforms like Rydberg atoms, quantum gases in optical cavities, and trapped ions serving as key tools for studying these interactions~\cite{Rempe2000, Saffman2010, Yao2021}. In such systems, noise both intrinsic and externally induced plays a critical role in determining the dynamics and behavior of quantum states. Noise effects the coherence and stability of quantum states, making it a key factor to consider when investigating long-range interactions. Understanding how noise influences quantum transitions, particularly in driven systems with long-range coupling, is essential for improving the robustness and efficiency of quantum simulations and computing.

Long-range models with power-law interactions $(J_{ij}\propto 1/|i-j|^{\alpha})$, have been widely explored in non-equilibrium quantum dynamics~\cite{Schachenmayer2013, Jaschke_2017,FerencIgloi2018, Uhrich2020,delCampo2022, NicoloDefenu2023}. One notable phenomenon is the quantum Kibble-Zurek mechanism (KZM), which describes the behavior of defects formed when a quantum system is adiabatically driven across a quantum critical point (QCP). In this process, the density of defects follows a universal scaling law, capturing fundamental aspects of quantum criticality~\cite{Zurek1985,ZUREK1996, Dziarmaga2010,Kolodrubetz2012,delcampo2014}. Recent studies have highlighted environmental effects on LR Kitaev and LR Ising models, offering deeper insight into their dynamics.~\cite{RicardoPuebla2020, AClerk2023,Kastner2023, FedericoCarollo2024,Langari2024}.
While much of the research has focused on the limiting cases involving $\alpha$, for example, the nearest neighbor transverse-field Ising model (TFIM) retrieved in the limit $\alpha\rightarrow \infty$~\cite{Dziarmaga2010,Kolodrubetz2012,delcampo2014}, and the Lipkin-Meshkov-Glick (LMG) model, which describes an all-to-all interaction or fully connected spin system in the limit $\alpha\rightarrow 0$~\cite{RosarioFazio2008,Acevedo2014,Defenu2018}, the effects of intermediate LR interactions on defect density, spin correlations, and their interplay with noise remains largely unexplored. Our current work aims to address this gap.

We initialize the long-range transverse-field Ising model (LR TFIM) in its ground state and adiabatically drive the transverse magnetic field. By introducing noise in the transverse term of the corresponding fermionic Hamiltonian, we explore the influence of LR interactions under noisy conditions. Our investigation encompasses defect density, spin correlations, and the full counting statistics of defects.
 We find that in the noiseless scenario, LR interactions enhance the defect density, while in the presence of noise increasing the range of interactions tends to suppress defect generation. Notably, in our driving protocol, the system crosses two QCPs, unlike the previous study~\cite{Anirban_dutta,N_defenu_TS}, where only one QCP was traversed. As a result, in the noiseless case, we observe that increasing long-rangedness leads to a higher defect density, opposite to the trend reported in~\cite{Anirban_dutta,N_defenu_TS}.
 This behavior is reflected in the longitudinal spin correlation function. Beyond the longitudinal spin correlation, we analyze various correlation functions, categorizing them into two types: type-I and type-II. Type-I correlations are expressed in terms of local fermionic two-point correlators, while type-II correlators typically involve string of fermionic correlators. Our findings reveal that type-I correlators exhibit an initial Gaussian decay followed by suppression, whereas type-II spin correlations decay exponentially. Another aspect we have investigated includes the statistics of the kink distribution. Recent studies have shown that the kink-number distribution, after the end of the drive protocol, exhibits universality beyond the KZM in SR models such as the one-dimensional TFIM, the XY model, and the Kitaev chain~\cite{delCampo2018,delcampo2021,King2022,MSingh_2023}. In the thermodynamic limit, defects produced by KZM follow a normal distribution in addition the higher cumulants of the kink-number distribution are proportional to the mean defect density and obey the universal KZ power-law scaling. However, in the recent work by Gherardini et al.~\cite{Defenu2024} demonstrated that the LMG model,  despite not obeying the Kibble-Zurek scaling law, exhibits universal behavior, in both the defect density and higher-order cumulants, with the defect statistics described by a negative binomial distribution.
Our work further investigates the previously unexplored influence of intermediate LR interactions and noise on these higher cumulants.  

 The paper is organized as follows. In Sec. II, we introduce the model Hamiltonian and the drive protocol. Sec. III presents the derivation of the Landau-Zener (LZ) transition probability and defect density for both noisy and noiseless scenarios. In Sec. IV, we analyze different types of spin correlators and investigate the effects of LR interactions and noise. Sec. V focuses on the full counting statistics (FCS) of defects and transverse magnetization. Finally, in Sec. VI, we summarize our findings, discuss potential future directions, and explore the possibilities for experimental validation.

\section{Model Hamiltonian and Driving protocol}
\begin{figure}[h]
    \centering
    \includegraphics[width=0.95\columnwidth,height=5.5cm]{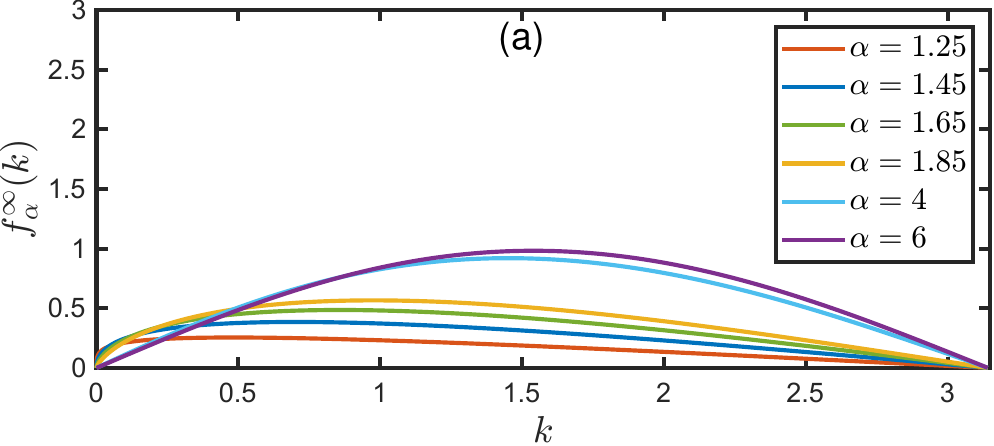}
    \includegraphics[width=0.95
    \columnwidth,height=4.75cm]{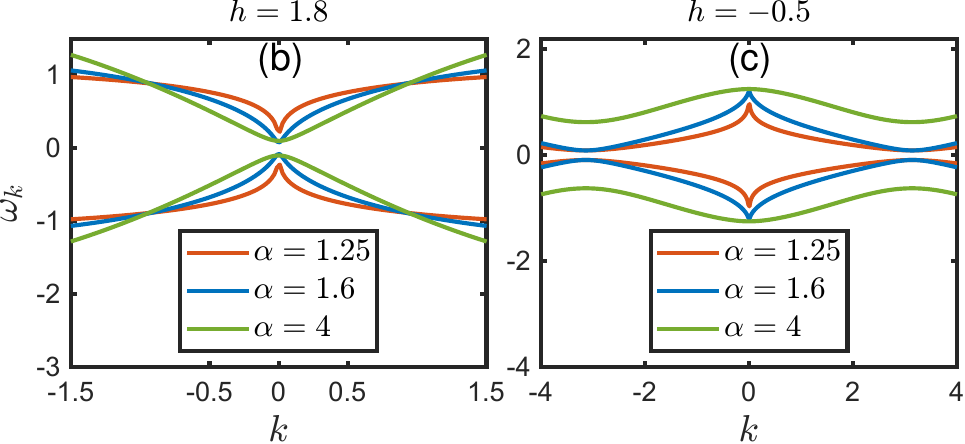}
    \caption{(a) The LR pairing term as a function of $k$ is plotted for different LR exponents. The plot reveals that the maximum shifts toward $k=0$ for $\alpha\leq 2$, while for $\alpha>2$ the maximum is closer to $k=\frac{\pi}{2}$. Panels (b) and (c) present the energy gap as a function of momentum $k$ for different values of $\alpha$ near the critical point, highlighting its behavior at $k=0$ for $h=1.8$ in Fig. (b) and at $k=\pi$ for $h=-0.5$ in Fig.(c).}
    \label{fig:falpha}
\end{figure}

We consider a one-dimensional LR Ising model, which includes nearest neighbor interaction and all possible cluster terms exhibiting interaction with a power law decay behavior~\cite{SchmidtKaiPhillip2016, DSadhukhan_2020,delCampo2022, Aditi_sen_LR_model, ChengxiangDing_2024}. For a system of $N$ spins, with $l=N-1$, number of clusters, the Hamiltonian of this model is defined as
\begin{equation}
\label{lr_hamiltonian}
    H=\sum_{n=1}^{N}[\frac{h}{2}\sigma_{z}^{n}+\sum_{r=1}^{l}J_{r}\sigma_{x}^{n}\prod_{i=n+1}^{n+r-1}\sigma_{z}^{i}\sigma_{x}^{n+r}],
\end{equation}
where,  $h$ is the transverse magnetic field, and $J_{r}=\frac{1}{\zeta(\alpha)r^\alpha}$, represents the coupling between the spin $\sigma^{i}_{x}$ and $\sigma^{j}_{x}$ with a normalization factor given by the Riemann
zeta function $\zeta(\alpha)= \sum_{r}\frac{1}{r^\alpha}$ (we consider $\alpha>1$). For $0<\alpha<1$, this model lies in the strongly interacting LR regime, and it belongs to the same class as the LMG model, while for $1<\alpha<2$, it belongs to the weakly interacting LR regime. For the case of $\alpha>2$, the Hamiltonian exhibits SR interacting behavior similar to the nearest-neighbor TFIM~\cite{DSadhukhan_2020,Aditi_sen_LR_model}. Our present work focuses on the regime, $1<\alpha<\infty$, and specifically refers to the range $1<\alpha<2$, as the LR regime throughout the paper.
We express the spin-operators in terms of the Fermionic operators via  the Jordan-Wigner (JW) transformation, 
\begin{equation}
    \begin{split}
        \sigma^{n}_{z} &=(1-2c^{\dagger}_{n}c_{n}),\\
        \sigma^{n}_{x} &=(c^{\dagger}_{n}+c_{n})\prod_{j<n}(1-2c^{\dagger}_{j}c_{j}).
    \end{split}
\end{equation}
The above spin model Eq.(\ref{lr_hamiltonian}) is mapped to a free fermionic Kitaev model with  LR pairing and hopping terms and is expressed as,
\begin{equation}
\begin{split}
\label{freeHamiltonian}
      H &=\sum_{n}\frac{h}{2}(1-2c^{\dagger}_{n}c_{n})+\sum_{n,r}J_{r}(c^{\dagger}_{n}c_{n+r}+c^{\dagger}_{n}c^{\dagger}_{n+r}-h.c.).\\
\end{split}
\end{equation}
A similar Kitaev model can be obtained from the LR quantum Ising model without clustering terms by performing a truncated JW transformation~\cite{Jaschke_2017,N_defenu_LR_2019}.
In the thermodynamic limit, $N\rightarrow \infty$,  Eq.~(\ref{freeHamiltonian}) acquires the following form  in  the momentum  space 
\begin{equation}
\begin{split}
      H &=\sum_{k>0}\Psi^{\dagger}_{k}\mathcal{H}^{0}_{k}\Psi_{k},
\end{split}
\end{equation}
where,
\begin{equation}
   \begin{split}
   \mathcal{H}^{0}_{k} &=2\begin{bmatrix}
h/2-g^{\infty}_{\alpha}(k) & i\,f^{\infty}_{\alpha}(k)\\
-i\,f^{\infty}_{\alpha}(k) & -h+g^{\infty}_{\alpha}(k)
\end{bmatrix}\\
&=2(h/2-g^{\infty}_{\alpha})\sigma_{z}-2f^{\infty}_{\alpha}\sigma_{y},
\end{split}
\end{equation}
 and  $\Psi_{k}=[c_{k},\,  c^{\dagger}_{-k}]^T$. The $k$ dependent terms   $g^{\infty}_{\alpha}(k)$, and $f^{\infty}_{\alpha}(k)$ are given by
\begin{align}
    g^{\infty}_{\alpha}(k)&= \sum_{r=1}^{}J_{r}\cos{kr}= \frac{1}{2\zeta(\alpha)}[\text{Li}_{\alpha}(e^{ik})+\text{Li}_{\alpha}(e^{-ik})]\nonumber\\
   f^{\infty}_{\alpha}(k)&=\sum_{r=1}^{}J_{r}\sin{kr} =\frac{1}{2i\zeta(\alpha)}[\text{Li}_{\alpha}(e^{ik})-\text{Li}_{\alpha}(e^{-ik})],\nonumber
\end{align}
where $\text{Li}_{\alpha}(x)=\sum_{l=1}^{\infty}\frac{x^l}{l^\alpha}$ and 
the eigenenergies are given by,
\begin{equation}
    \omega_{k}=\pm2\sqrt{(h/2-g^{\infty}_{\alpha})^2+(f^{\infty}_{\alpha})^2}.
\end{equation}
As shown in Fig.~(\ref{fig:falpha}), LR interaction drastically affects the pairing term associated with the energy gap. 
Unlike the SR regime where,  $f^{\infty}_{\alpha}(k)$ exhibits symmetry about $k=\pi/2$,  in the LR regime it becomes asymmetric. 
 This asymmetric behavior of $f^{\infty}_{\alpha}(k)$ affects the Landau Zener (LZ) transition probability for both the noiseless and noisy drive scenarios when driven across QCP.
 The energy gap closes at the QCPs associated with the $k_c=0$, and $k_c=\pi$ modes. The mode, $k_c=0$ corresponds to the critical point, $h_{c}^{1}=2$, while for the mode $k_c=\pi$, the critical point is given by $h_{c}^{2}=\frac{2}{\zeta(\alpha)}\sum_{r}\frac{(-1)^r}{r^\alpha}=(2^{1-\alpha}-1)$. Note that in the SR regime, $h^{2}_{c}\rightarrow-1$. 
The region between $h_{c}^{1}$ and $h_{c}^{2}$ belongs to the ordered phase (ferromagnetic phase), while the rest of the region is in the disordered phase (paramagnetic phase) \cite{DSadhukhan_2020,dziarmaga_2020,Aditi_sen_LR_model,ChengxiangDing_2024}.

We consider a driving protocol where the transverse magnetic field, $h(t)$, is varied linearly with a driving rate, ${\tau_Q}$, i.e., $h(t)=\frac{t}{\tau_{Q}}$, from the paramagnetic phase ($t=-\infty$) to the paramagnetic phase ($t=\infty$). For each $k$ mode, $\mathcal{H}^{0}_{k}(t)$ is mapped to an independent LZ transition problem.
Next, we consider the case where a time-dependent noise is present in the off-diagonal term of the momentum-space Hamiltonian. Under this noisy driving scenario, the momentum-space Hamiltonian can be expressed as
\begin{equation}
\begin{split}
      \mathcal{H}^{\eta_0}_k&=\mathcal{H}^{0}_{k}(t)-\eta_{}(t)f^{\infty}_{\alpha}(k)\sigma_{y}\\
      &=E_{k}(t)\sigma_{z}+\bar{\Delta}_{k}\sigma_{y},
      \label{noisyHamiltonian}
\end{split}  
\end{equation}
 where $E_{k}(t)=2(h/2-g^{\infty}_{\alpha})$, $\bar{\Delta}_{k}=-2(f^{\infty}_{\alpha}+\eta_{}(t)f^{\infty}_{\alpha})$.  The term $\eta(t)$ represents time-dependent fluctuation characterized by the noise correlation function,  $\langle \eta(t)\eta(t_1) \rangle = \eta_{0}^2 e^{-\gamma |t-t_1|}$, where $\eta_{0}$ denotes the noise amplitude and $1/\gamma$ defines the noise correlation time. We consider the limit, $\gamma\rightarrow\infty$, in which the noise becomes delta-correlated, corresponding to white Gaussian noise.

\section{ Defect density scaling for noiseless and noisy drive}
In the momentum-space representation, the system is initially prepared in the ground state $\ket{0_k}$ which corresponds to the density matrix $\rho_{k}(0)=\ket{0_k}\bra{0_k}$. The dynamics of the system is governed
by the von Neumann equation,
\begin{equation}
\label{VnE}
    \frac{d\rho_{k}}{dt}=-i[\mathcal{H}^{}_{k}, \rho_{k}],
\end{equation}
where $\mathcal{H}_k$ is the Hamiltonian of the system and preserves the parity of the fermion occupancy number for both the noisy and noiseless cases. Since the ground state $\ket{0_k}$ and the excited state $\ket{k,-k}$ belong to the even-parity subspace, the  dynamics remain confined to this subspace. The total density matrix of the system is given by, $\rho=\otimes_{k}\rho_{k}$. 
 At long times, the diagonal components of $\rho_{k}$ become smooth functions of $k$, while the off-diagonal terms oscillate rapidly.
 Consequently, the momentum integrals involving the off-diagonal term, for example the two-point fermionic correlators $(\langle c^{\dagger}_{x}c_{x+r}^{\dagger} \rangle,\langle c^{}_{x}c^{}_{x+r} \rangle)$,  vanishes for any $r$.
 Any observable that can be expressed in terms of the two-point correlators will have no contribution from the off-diagonal terms of the density matrix. Therefore, neglecting the off-diagonal elements of $\rho_{k}$, we can infer that as $t\rightarrow \infty$, the system evolves into a nonequilibrium steady state (NESS), and is effectively represented by a decohered density matrix,
\begin{equation}
    \rho_{k}^{s}= p^{}_{k}\ket{0_k}\bra{0_k}+(1- p^{}_{k})\ket{k,-k}\bra{k,-k}
\end{equation}
where, $p_{k}$ is the LZ transition probability~\cite{Levitov_2006,MSingh_2021}. Next, we analyze $p_{k}$ and defect density to study the effect of LR interaction for both the noisy and noiseless scenario. The defect density is determined by integrating $p_{k}$ over the momentum space.  
\begin{figure}[h]
    \centering
    \includegraphics[width=0.9\columnwidth,height=5cm]{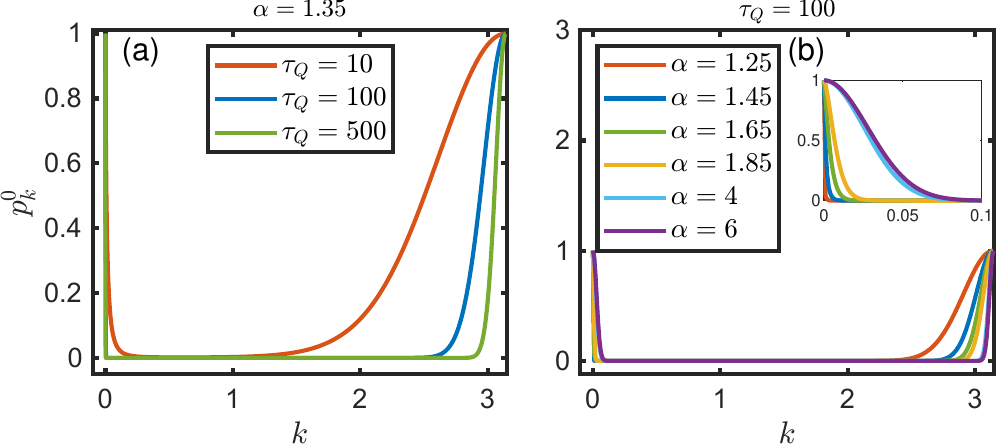}
    \caption{The Landau-Zener transition probability has been plotted as a function of momentum in the absence of noise. Figures (a) and (b) illustrate the dependence on quench time $\tau_Q$ and LR decay exponent $\alpha$ respectively. In Figure (b), the inset highlights the behavior of $p_k$ near $k=0$. }
    \label{fig:noiseless_pk}
\end{figure}

\subsection{Noiseless Drive}
\begin{figure}[h]
    \centering
    \includegraphics[width=\columnwidth,height=5.6cm]{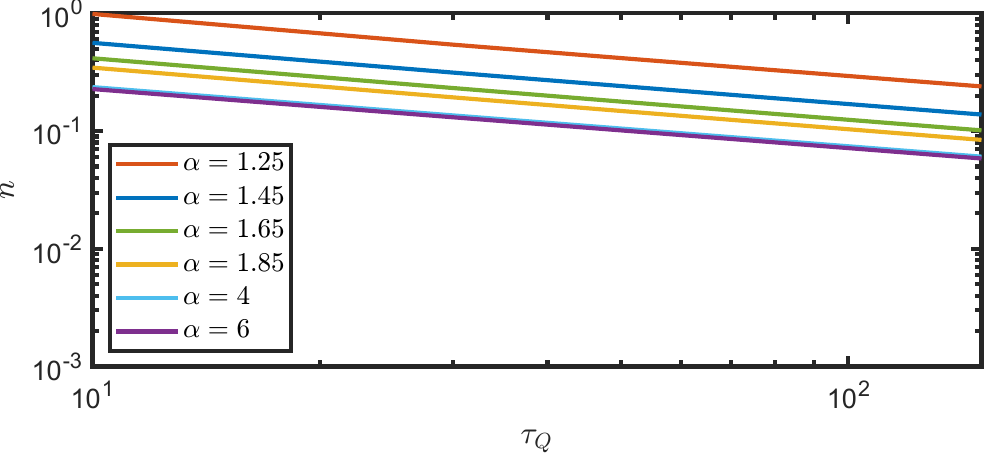}
    \caption{The plot shows the defect density as a function of quench time $(\tau_Q)$ for different values of the LR decay exponent $(\alpha)$ in the absence of noise.}
    \label{fig:noiseless_dd}
\end{figure}

For the noiseless case, the LZ transition probability (survival probability) is given by,
\begin{equation}
     p^{0}_{k} =e^{-2\pi\tau_{Q}(f^{\infty}_{\alpha})^2}.
\end{equation}
For $1<\alpha \leq 2$, $p^{0}_{k}$ can be approximated as
\begin{equation}
    \begin{split}
        p^{0}_{k} &=e^{-2\pi\tau_{Q}(f^{\infty}_{\alpha})^2}\\
        &\approx e^{-\chi\tau_{Q}k^\gamma}+e^{-\phi\tau_{Q}(\pi-k)^2},
    \end{split}
\end{equation}
where $\chi(\alpha)=\pi  [\cos{\frac{\pi\alpha}{2}}\frac{\Gamma(1-\alpha)} {\zeta{(\alpha)}}]^2$ and $\phi(\alpha)=\pi[\frac{\zeta(\alpha-1)}{\zeta(\alpha)}]^2$, where the parameter $\gamma$  is given by $\gamma=2(\alpha-1)$. The approximate expression for $p_{k}^0$ is obtained by expanding $f^{\infty}_{\alpha}(k)$ around the $k_c=0$ and $k_c=\pi$ regions (see Appendix: A).
The defect density is obtained by integrating  $p_{k}^0$ over the $k$ modes,
 \begin{equation}
 \label{noselessdefectdensity}
 \begin{split}
        n^{}_{}&=\frac{1}{\pi}\int_{0}^{\pi}dk p^{0}_{k}\\
        &\approx A(\alpha)(\tau_{Q})^{-1/\gamma}+B(\alpha)(\tau_{Q})^{-1/2},
 \end{split}
 \end{equation}
 where, $A(\alpha)=\frac{1}{\pi\gamma}\chi^{-1/\gamma}\Gamma \Bigl(1+\frac{1}{\gamma}\Bigl)$ and $B(\alpha)=\frac{1}{2}(\frac{\phi}{\pi} )^{-1/2}$.
 The first term in Eq.~(\ref{noselessdefectdensity}) accounts for the defect density arising from the modes near $k_c=0$, while the second term corresponds to the contributions from the modes about $k_c=\pi$. The region around these points are progressively suppressed as $\tau_Q$ is increased.   In the slow drive regime, the second term gives the dominant contribution to the defect production, thus implying that the defect density exhibits the expected universal KZ power-law scaling behavior, $n_{}\propto{\tau_Q^{-1/2}}$. The negligible contribution due to the first term can be understood from the energy gap structure at $k_c=0$, which closes sharply for the LR regime [see Fig.\ref{fig:falpha}(b)]. In contrast, as shown in Fig.\ref{fig:falpha}(c) near $k_c=\pi$, the gap closes more gradually as the range of LR interactions increase, leading to most of the defect production being driven by modes around the $k_c=\pi$ regions and resulting in enhanced defect production with the increase of range of interaction [see Fig.(\ref{fig:noiseless_dd})]. The scaling of defect density with the quench time and enhancement of its magnitude with decreasing $\alpha$, differs from the recent studies~\cite{Anirban_dutta,N_defenu_TS} on quench dynamics in the long-range Kitaev model, where the system undergoes a transition from the trivial to the topological phase.
 
For $\alpha >2$, i.e., the SR regime, equivalent number of modes are excited from both the $k=0$ and $k=\pi$ regions, yielding KZ power-law scaling behavior, $n=2B(\alpha)\,\tau_{Q}^{-1/2}$. As $\alpha\rightarrow\infty$, the defect density becomes independent of the $\alpha$.  
\begin{figure}[h]
    \centering
    \includegraphics[width=\columnwidth,height=5.5cm]{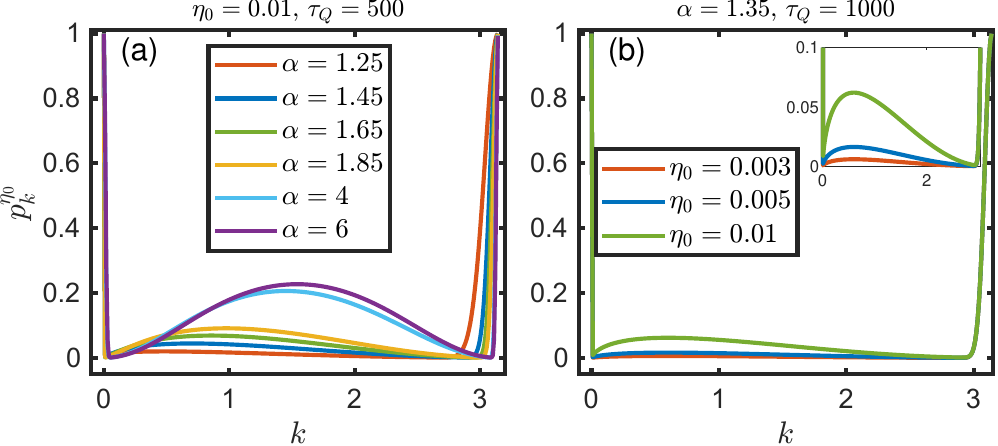}
    \caption{The LZ transition probability in the presence of noise is plotted as a function of momentum modes in Figures (a) and (b), corresponding to different $\alpha$ values and noise strengths, respectively. The inset in Figure (b) illustrates the effect of noise on the LZ transition probability for a fixed value of decay exponent $\alpha=1.35$. Both Figures (a) and (b) exhibit asymmetry about $k=\pi/2$ in the LR case and show a clear deviation from SR models. }
    \label{fig:noisypk}
\end{figure}
\subsection{Noisy Drive}
\begin{figure}[h]
    \centering
    \includegraphics[width=\columnwidth,height=8.5cm]{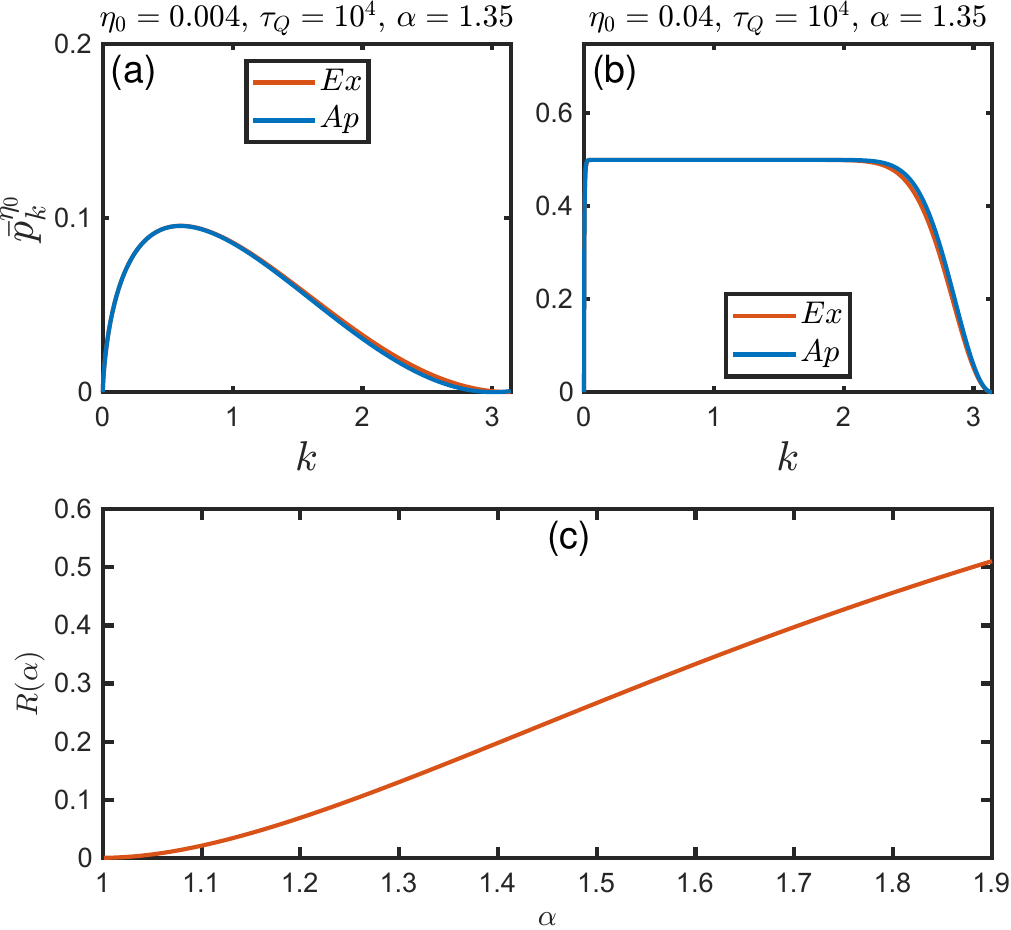}
    \caption{The comparison between the exact, $Ex$ and the approximation,  $Ap$ of $\bar{p}_{k}^{\eta_0}$ has been depicted in the LR regime $(\alpha=1.35)$. Figure (a) corresponds to the regime $\eta_{0}^2\tau_Q\ll1$, whereas figure (b) represents the case $\eta_{0}^2\tau_Q\gg1$. Fig.(c), illustrates the behavior of $R(\alpha)$ as a function of $\alpha$, revealing that as $\alpha$ increases, $R(\alpha)$ grows accordingly. }
    \label{fig:noisypkapprox}
\end{figure}

In the noisy drive scenario, the LZ transition probability is given by (see Appendix: B),
\begin{equation}
       p^{\eta_0}_{k}=\frac{1}{2}[1+e^{-4\pi\tau_{Q}\eta_{0}^2(f^{\infty}_{\alpha})^2}(2p^{0}_{k}-1)]
\end{equation}
In the slow drive regime and $\eta_0^2 \ll 1$,  $p^{\eta_0}_{k}$ can be approximated as $p^{\eta_0}_{k}\approx p^{0}_{k}+\bar{p}^{\eta_0}_{k}$, i.e., a noiseless part and an additional part due to noise. In contrast, to the noiseless case, contributions from the noise arise predominantly from modes away from the above gap closing points. While for the SR case, the important modes are around $k\approx \pi/2$, for the LR case they are progressively left shifted [see the Fig.(\ref{fig:noisypk})]. This shift is accompanied by a pronounced suppression of noise-induced excitations as either $\alpha$ or the noise amplitude is reduced, as shown in Fig.(\ref{fig:noisypk})(a) and (b) respectively.
For the defect density the contribution from the noisy part $\bar{p}^{\eta_0}_{k}$ dominates in either of the two regimes, $\eta_{0}^2\tau_{Q}\ll1$ and $\eta_{0}^2\tau_{Q}\gg1$. 
For $\eta_{0}^2\tau_{Q}\ll1$, $\bar{p}^{\eta_0}_{k}$ can be approximated as
\begin{equation}
\begin{split}
      \bar{p}_{k}^{\eta_0}&\approx \frac{1}{2}[1-e^{-4\tau_{Q}\eta_{0}^2 F(k,\alpha)}],
\end{split}
\end{equation}
where, $F$ is obtained by expanding $(f^{\infty}_{\alpha})^2$ about $k=0$, and $F(k,\alpha)=[\chi k^\gamma+\phi k^2+d(\alpha)k^{\alpha}]$ with $d(\alpha)=2\pi\cos{\frac{\pi\alpha}{2}}\frac{\Gamma(1-\alpha)}{\zeta{(\alpha)}}\frac{\zeta(\alpha-1)}{\zeta(\alpha)}$. The comparison between the exact and approximation is shown in Fig.\ref{fig:noisypkapprox}(a).

 Furthermore, $\bar{p}_{k}^{\eta_0}$ can be expressed as $\bar{p}_{k}^{\eta_0}\approx 2\eta_{0}^{2}\tau_{Q}[\chi k^\gamma+\phi k^2+d(\alpha)k^{\alpha}]$, and the contribution to
defect density from the noise is given by
\begin{equation}
\label{noisydd1}
\begin{split}
       n_{}&\approx \frac{1}{\pi} \int_{0}^{\pi} dk\, 2\eta_{0}^{2}\tau_{Q}[\chi k^\gamma+\phi k^2+d(\alpha)k^{\alpha}]\\             
       &= \eta_{0}^{2}\tau_{Q}R(\alpha),
\end{split}
\end{equation}
where, $R(\alpha)$ is a prefactor that depends on the LR exponent and its behavior is shown in Fig.\ref{fig:noisypkapprox}(c). 

\begin{figure}
    \centering
    \includegraphics[width=\columnwidth,height=12cm]{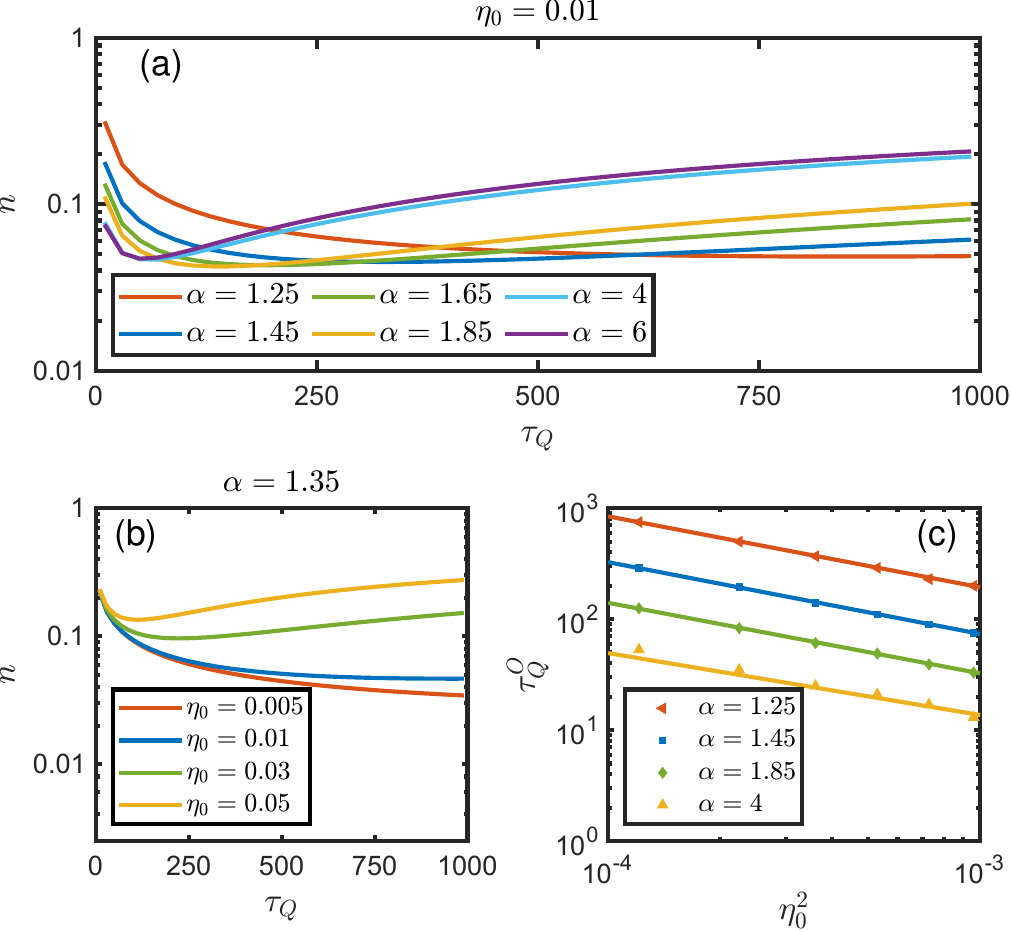}
    \caption{The interplay of noise and  LR exponents in the defect density have been shown where defect density is calculated using the exact expression of $p_{k}^{\eta_0}$. Figure.(a) shows the scaling of defect density $n$ with quench time $\tau_Q$ for different values of $\alpha$ at a fixed noise strength $\eta_{0}=0.01$. Whereas in Figure (b)  shows the defect density scaling for various noise strengths, keeping the decay exponent fixed at $\alpha=1.35$. Figure.(c) depicts the optimal quench time $(\tau_Q^{O})$ as a function of the square of noise amplitude ($\eta_{0}^2$) for different $\alpha$. The plot exhibits the universal scaling of the optimal quench time.}
    \label{fig:noisy_defect_density}
\end{figure}

In the regime where, $\eta_{0}^2\tau_{Q}\gg1$, $\bar{p}_{k}^{\eta_0}$ can be approximated as follows,
\begin{equation}
    \begin{split}
        \bar{p}_{k}^{\eta_0}\approx\frac{1}{2}[1-e^{-4\eta_{0}^{2}\tau_{Q}\chi k^{\gamma}}-e^{-4\eta_{0}^{2}\tau_{Q}\phi (\pi-k)^2 }].
    \end{split}
\end{equation}
We compared the exact and approximation in Fig.\ref{fig:noisypkapprox}(b).
 In the large slow drive regime, the contribution of noise to the defect density is given by
\begin{equation}
\label{noisydd2}
\begin{split}
       n_{} &
       \approx \frac{1}{2\pi} \int_{0}^{\pi} dk\,[1-e^{-4\eta_{0}^{2}\tau_{Q}\chi k^{\gamma}}-e^{-4\eta_{0}^{2}\tau_{Q}\phi (\pi-k)^2 }]\\
       &\approx \frac{1}{2}-\frac{A(\alpha)}{2}(4\tau_{Q}\eta_{0}^2)^{-1/\gamma}-\frac{B(\alpha)}{2}(4\tau_{Q}\eta_{0}^2)^{-1/2}.
\end{split}
\end{equation}
Eq.(\ref{noisydd2}) indicates a suppression in defect density, with the last term in Eq.(\ref{noisydd2}) dominating this effect in the limit $\eta_{0}^2\tau_{Q}\gg1$. Consequently, the suppression becomes more pronounced as $\alpha$ decreases, causing the defect density to approach its maximum more gradually.
The analytical results reveal that for $\eta_{0}^2\tau_Q\ll1$,  the defect density increases with both the driving rate and noise strength, indicating the anti-Kibble-Zurek (AKZ) behavior. However, as the range of LR interaction is increased, the defect density is suppressed. 
These findings are further supported by numerical results obtained using the exact expression for $p_{k}^{\eta_0}$, as shown in Fig.(\ref{fig:noisy_defect_density}) and,  we observe that in the noisy scenario, defect density is suppressed with the decrease of $\alpha$, and noise strength $\eta_0$ [see the Figs~\ref{fig:noisy_defect_density}(a)-(b)]. 
Since noise couples to the off-diagonal elements of the Hamiltonian for each $k$ mode as described in Eq.(\ref{noisyHamiltonian}), its impact is influenced by the structure of $f^{\infty}_{\alpha}(k)$. As $\alpha$ decreases, the magnitude of $f^{\infty}_{\alpha}(k)$ reduces [see the Fig.\ref{fig:falpha}(a)], and results in suppression of the noise induced effect which is manifested in both the LZ transition probability and defect density.
\\

In the LR regime, as shown in Fig.~\ref{fig:noisy_defect_density}(c) we observe that the optimal quench time $(\tau_Q^{O})$ follows the universal scaling law, $\tau_{Q}^{O}\propto \eta_{0}^{-4/3}$, similar to the SR model. However, due to the suppression of defect density with the decrease of $\alpha$ under noisy conditions, the defect density reaches the minima for larger quench time and results in the shifting of the optimal quench time with $\alpha$.  Specifically, $\tau_Q^{O}$ decreases as the value of $\alpha$ grows, highlighting the influence of interaction range on optimal driving time.

\section{Interplay between noise and LR interaction in spin correlation}

In the framework of KZM, studies on SR models, such as the TFIM and the XY model, have shown that in the asymptotic limit, the longitudinal spin-spin correlator  $(\langle \sigma_{x/y}^{n}\sigma_{x/y}^{n+r} \rangle)$, decays exponentially with the spatial separation, whereas the transverse correlator $(\langle \sigma_{z}^{n}\sigma_{z}^{n+r} \rangle)$ follows a spatial Gaussian decay~\cite{Levitov_2006, Zurek2007}. Interestingly, recent works have also revealed that the correlation between KZM-generated defects exhibits Gaussian decay similar to the transverse correlator~\cite{KRC_2021,MarekRams_2022} while noise enhances the correlation length of the transverse correlator~\cite{MSingh_2021}.

Following the convention of Ref.~\cite{Levitov_2006}, we express the spin operators in terms of the Majorana fermionic operators, $A_{x}=c^{\dagger}_{x}+c_{x}$ and $ B_{x}=c^{\dagger}_{x}-c_{x}$ where $c_{x}$  and $c^{\dagger}_x$ are the fermionic creation and annihilation operators. In the final decohered state, only the cross correlators $G(n-n')=\langle A_{n}B_{n'}\rangle$ remain nonzero while the correlators of the type, $ \langle A_{n}A_{n'}\rangle,  $ and $\langle B_{n}B_{n'}\rangle$ vanish. Since the Wick's theorem holds, the spin correlators can be expressed the products of two-point correlator of the type $\langle A_{n}B_{n'}\rangle$. Before analyzing the spin correlators, we study the behavior of two-point Majorana correlators for both the noisy and noiseless scenarios.

\begin{figure}
    \centering
    \includegraphics[width=\columnwidth,height=5.5cm]{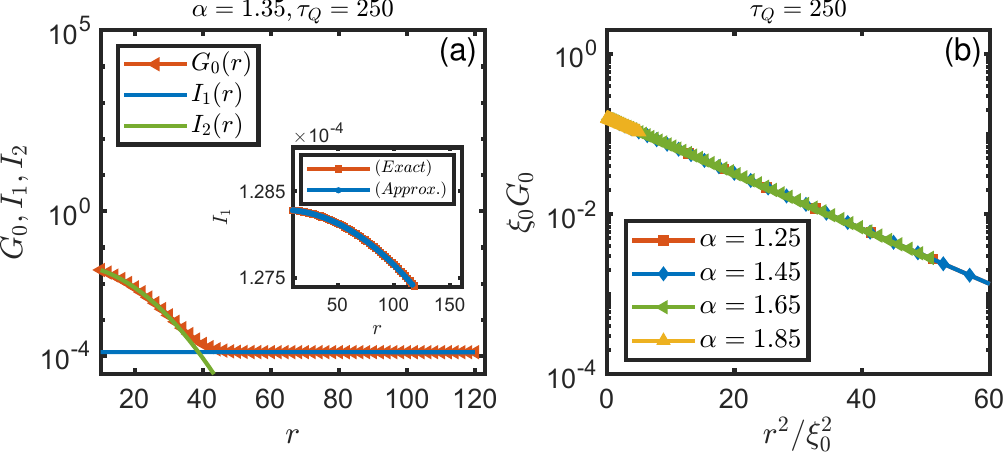}
    \caption{The two-point correlation $G_{0}(r)$ is plotted as a function of separation $r$ in the long-range (LR) regime. Figure (a) demonstrates that  $G_{0}(r)$  can be approximated by $I_1$  and $I_2$ for $\alpha = 1.35$ and quench time $\tau_Q = 250$, where $I_2$ corresponds to a Gaussian decay, while $I_1$ represents a nontrivial decay. The inset in Figure (a) compares the exact and approximate forms of $I_1$. Figure (b) illustrates the Gaussian decay behavior for different $\alpha$ values at $\tau_Q = 250$.
}
    \label{fig:noiseless_2point_correlator}
\end{figure}

For the noiseless scenario two-point correlator is given by,
\begin{equation}
\label{twopointcorlator}
    \begin{split}
        G_{0}(r) =\frac{1}{2\pi}\int_{-\pi}^{\pi}dk\, e^{i k r}\, p^{0}_{k},
    \end{split}
\end{equation}
where $p_k^0$ corresponds to the noiseless LR regime. It  can be expressed in two parts, $G_{0}(r)= I_{1}(r)+ I_{2}(r)$ with,
\begin{equation}
     \begin{split}
     I_{1}(r,\tau_Q)&=\frac{1}{\pi}\int_{0}^{\pi}dk\, \cos{kr}\, e^{-\chi\tau_{Q}k^\gamma},\\
     &\approx A(\alpha)\tau_{Q}^{-1/\gamma} - \frac{r^2}{2\gamma \chi^{3/\gamma}} \Gamma\left(\frac{3}{\gamma}\right)\tau_{Q}^{-3/\gamma}\\
     I_{2}(r,\tau_Q)&=\frac{1}{\pi}\int_{0}^{\pi}dk\, \cos{kr}\,e^{-\phi\tau_Q(\pi-k)^2}\\
     &\approx \frac{e^{-\frac{r^2}{4\phi \tau_{Q}}}}{\sqrt{4\pi\phi\tau_Q}},
     \end{split}
 \end{equation}
where, $I_1(r)$ is the contribution from the $k=0$ regions, while $I_{2}(r)$ originates from the $k=\pi$ region.
 At the end of protocol, i.e., deep in the paramagnetic phase, at the length scale $r^{*}\approx[-4\phi \tau_{Q}\, \text{ln}(\frac{A \sqrt{4\pi\phi\tau_Q} }{2}\, \tau_Q^{-1/\gamma+1/2})]^{1/2}$, and $I_{2}$ and $I_{1}$  are of similar order. For $r< r^{*}$, $I_2(r)$ dominates resulting in  Gaussian decay behavior with the correlation length $\xi_0=\sqrt{4\phi\tau_Q}$ while for $r\gg r^{*}$, $I_{1}$ dominates causing the correlator to suppress quadratically with the separation as shown in Fig.\ref{fig:noiseless_2point_correlator}(a). 
 In contrast, for $\frac{\chi\tau_Q}{r^{\gamma}}\ll1$  the two-point correlator transitions to an algebraic decay, scaling as $I_{1}\propto\frac{1}{r^{2\alpha-1}}$.
 We note that, in the short-range regime, $I_1$ exhibits Gaussian decay, leading to SR behavior of the two-point correlator across all length scales, consistent with earlier results.

The two-point correlator in the presence of noise is obtained by replacing $p_k^{0}$ in Eq.(\ref{twopointcorlator}) with  $p_k^{\eta_0}$
\begin{equation}
    \bar{G}(r)=\frac{1}{2\pi}\int_{-\pi}^{\pi}dk\, e^{i k r}\, p^{\eta_{0}}_{k}.
\end{equation}
Since the LZ probability $(p_{k}^{\eta_{0}})$ can be decomposed into contributions from both noisy and noiseless components, therefore, the two-point correlator can be expressed as the sum of these two contributions,
\begin{equation}
     \bar{G}(r)= G_{\eta_{0}}+G_{0},
\end{equation}
where $G_{\eta_0}$ accounts for the noise contribution, and $G_{0}$ represents the noiseless component.

\begin{figure}
    \centering
    \includegraphics[width=\columnwidth,height=5.5cm]{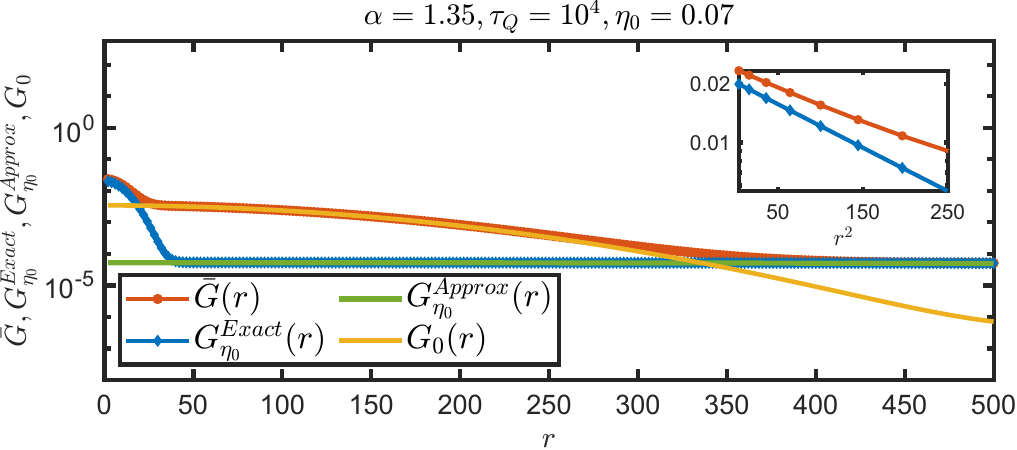}
    \caption{In the LR regime for $\eta_0^2\tau_Q\gg1$, the two-point correlator is plotted as a function of separation $r$ where we compare the full two-point correlator $\bar{G}(r)$, the exact noisy correlator $G^{Exact}_{\eta_0}$,  the approximate noisy correlator 
$G^{Approx}_{\eta_0}$ and the noiseless correlator $G_{0}$. The inset plot highlights the Gaussian decay with different correlation lengths, showing a transition from $G^{Exact}_{\eta_0}$ to  $G_{0}$. Where the first crossover in $\bar{G}(r)$ occurs at $\tilde{r}$, followed by another transition from
$G_{0}$ to $G^{Exact}_{\eta_0}$ at $\bar{r}$ . Here, we considered a fixed $\alpha=1.35$, quench time $\tau_Q=10^4$ and noise strength $\eta_{0}=0.07$. }
    \label{fig:noisy_2point_correlator}
\end{figure}

In the slow driving regime, where $(\eta_{0}^2\tau_Q\gg 1)$, and long distances, $r\gg 1$  (assuming $r$ is even), the noisy contribution to the two-point correlator, $G_{\eta_{0}}$, can be approximated as,
\begin{equation}
    \begin{split}
         G_{\eta_0} &= -\frac{1}{\pi}\int_{0}^{\pi}dk\, [e^{-4\chi\eta_{0}^2\tau_Q k^{\gamma}}+e^{-4\phi\eta_{0}^2\tau_Q (\pi-k)^{2}}]\cos{kr}\\
         &\approx -\frac{1}{\pi}[\int_{0}^{\pi}dk\, e^{-4\chi\eta_{0}^2\tau_Q k^{\gamma}} (1-k^2r^2/2)+\frac{e^{-\frac{r^2}{16\phi\eta_{0}^2\tau_Q}}}{\sqrt{16\pi\phi\eta_{0}^2\tau_Q}}]\\
          &=I_{1}(r,4\eta_0^2\tau_Q)+I_{2}(r,4\eta_0^2\tau_Q).
    \end{split}
\end{equation}
 It turns out that, unlike the noiseless scenario, the two-point correlator $\bar{G}$ exhibits twofold crossover. Initially, it undergoes a transition from a Gaussian decay with correlation length $\xi=\sqrt{16\phi\eta_{0}^2\tau_Q}$ to the $\xi=\sqrt{4\phi\tau_Q}$,  followed by an algebraic suppression governed by $G_{\eta_{0}}$ [see Fig.(\ref{fig:noisy_2point_correlator})]. The first crossover occurs at $\tilde{r}=[-16\phi \eta_0^2\tau_Q\text{ln}(2\eta_0)]$, while the second crossover is given by $\bar{r}\approx[-4\phi \tau_{Q}\, \text{ln}(\frac{A \sqrt{4\pi\phi\tau_Q} }{2}\, \eta_{0}^{-2/\gamma}\tau_Q^{-1/\gamma+1/2})]^{1/2}$.
 
Next, we explore the effects of LR interaction on the spin correlation in the NESS for noisy and noiseless scenarios and consider the spin-spin correlators defined as follows,

 \begin{equation}
     \begin{split}
         \mathcal{C}^{ll}(r)&=\langle \sigma_{l}^{n}\sigma_{l}^{n+r} \rangle- \langle \sigma_{l}^{n} \rangle \langle \sigma_{l}^{n+r} \rangle,\\
         \rho^{jj}(r) &=\langle(1-\mu_{j}^{n})(1-\mu_{j}^{n+r})\rangle-\langle (1-\mu_{j}^{n}) \rangle^2,\\
     \end{split}
 \end{equation}
where $\mu_{j}^{n}=\sigma_{j}^{n}\sigma_{j}^{n+1}$ with $l,j=x,z$, and string correlators,
 \begin{equation}
     \begin{split}
        \mathcal{A}(r)&=\langle\sigma_{x}^{n}\sigma_{x}^{n+r}\prod_{i=n+1}^{n+r-1}\sigma_{z}^{i}\rangle,\\
        \mathcal{E}(r)&=\langle \prod_{l}\frac{(1-\sigma_{z}^{l})}{2} \rangle.
     \end{split}
 \end{equation}
The string correlator, $\mathcal{A}(r)$ is associated with the spin-spin interaction term in the LR Hamiltonian defined in Eq.(\ref{lr_hamiltonian}), and has been a subject of extensive investigation in spin models
~\cite{TonyLee2016,DSadhukhan_2020,Maghrebi2024,Teretenkov2024,Mi2024}. On the other hand, $\mathcal{E}(r)$ referred to as the emptiness probability, quantifies the probability of observing $r$ adjacent spins in the down state in the final configuration at the end of the ramp protocol~\cite{Nishiyama2001,Abanov_Franchini_2005,Ares_2020,MarekRams_2022}.

\subsection{Type-I spin correlation}
As discussed earlier spin-correlation functions such as $\mathcal{C}^{zz},\, \rho^{zz}$, and $\mathcal{A}(r)$ belong to the type-I class, they can be expressed in terms of either the linear or quadratic powers of the two-point fermionic correlator  $G(r)$. Consequently, these spin correlators exhibit behavior closely aligned with that of $G(r)$. In the following, we explicitly analyze $\mathcal{A}(r)$. The exact expressions for other correlators in terms of two-point fermionic correlators are provided in Appendix C. The string correlator, $\mathcal{A}(r)$  becomes local in terms of Majorana operators and can be expressed as
\begin{equation}
\begin{split}
     \mathcal{A}(r)&=\langle \sum_{n}\sigma_{x}^{n}\sigma_{x}^{n+r}\prod_{i=n+1}^{n+r-1}\sigma_{z}^{i}\rangle\\
     &=\frac{1}{2} \sum_{n}\Big(\langle B_{n}A_{n+r} \rangle+ \text{h.c.}\Big)\\
      &=\frac{1}{2\pi}\int_{-\pi}^{\pi}dk\, (1-2p_{k})\cos{kr}.
\end{split}
\end{equation}
From the earlier results of two-point fermionic correlators, we deduce that at the end of the ramp the string correlator $\mathcal{A}(r)$ exhibits Gaussian decay for all values of $\alpha$. In contrast, recent investigations on the SR XY model by Paul et al., \cite{Maghrebi2024}  and on SR TFIM by Mi et al., \cite{Mi2024} 
wherein sudden quench  and 
 dissipative cooling protocols were examined, respectively,  reveal an exponential decay in the paramagnetic phase. 
 This difference has been attributed to the fact that eigenstates of the system cannot inherently generate Gaussian correlations~\cite{KRC_2021}.
Our results suggest that the nonequilibrium state induced by the linear ramp protocol and the linear dispersion near the QCP, $k=\pi$ is responsible for the emergence of Gaussian decay. Similarly, the correlators  $\mathcal{C}^{zz},\, \rho^{zz}$ exhibits the Gaussian decay behavior. The result is valid for both the noisy and noiseless scenario.

\subsection{Type-II spin correlation}
\begin{figure*}
    \centering
    \includegraphics[width=\columnwidth,height=4.25cm]{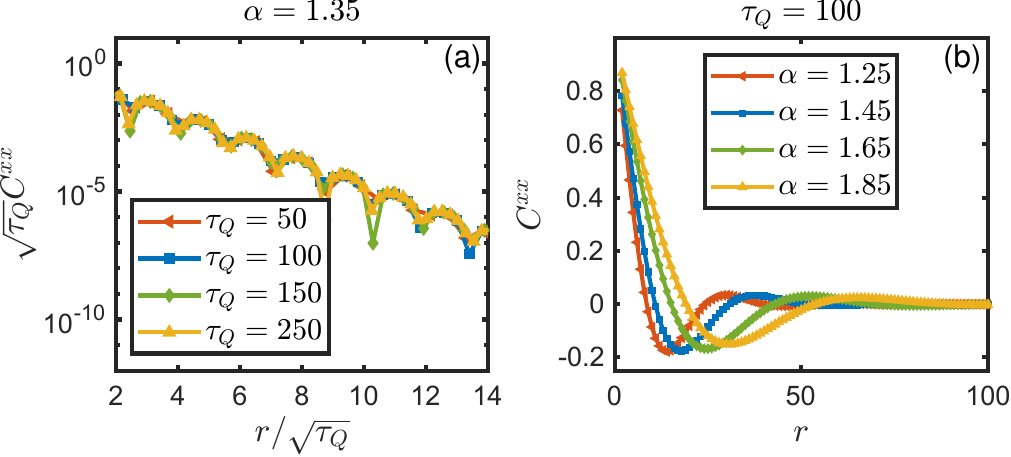}
    \includegraphics[width=\columnwidth,height=4.25cm]{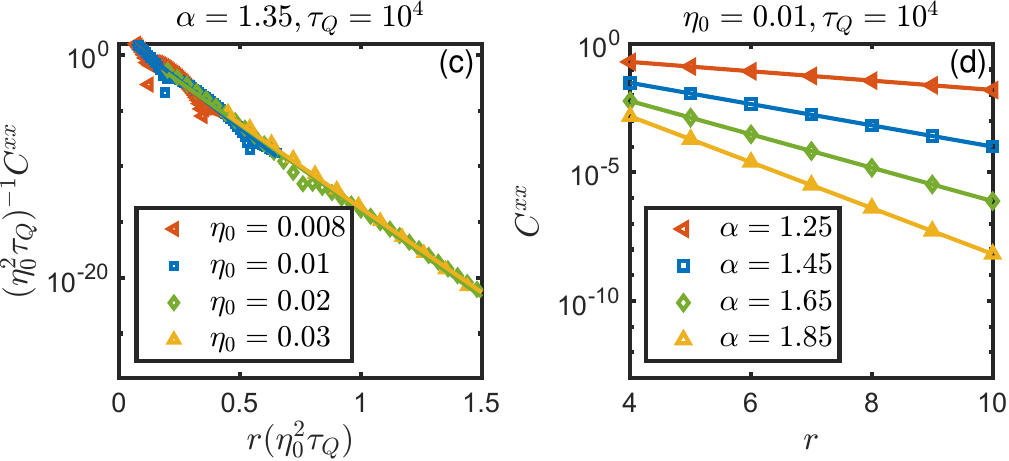}
    \caption{The rescaled longitudinal spin correlation $(\mathcal{C}^{xx})$ is plotted as a function of separation $(r)$ for both noisy and noiseless scenarios. In the noiseless scenario, fig.~$(a)$ illustrates the scaling of the spin correlator with quench time $\tau_Q$ for $\alpha = 1.35$, while fig.~\((b)\) shows its dependence on the LR interaction decaying parameter $\alpha$ for \(\tau_Q = 250\). In Figure \((c)\), the spin correlation is plotted for different noise strengths at a fixed  $(\alpha = 1.35)$, whereas fig.~\((d)\) represents its behavior for different $\alpha$ values in the noisy scenario with \(\eta_{0}= 0.04\).
}
    \label{fig:spinxx}
\end{figure*}
\begin{figure}
    \centering
    \includegraphics[width=\columnwidth,height=5cm]{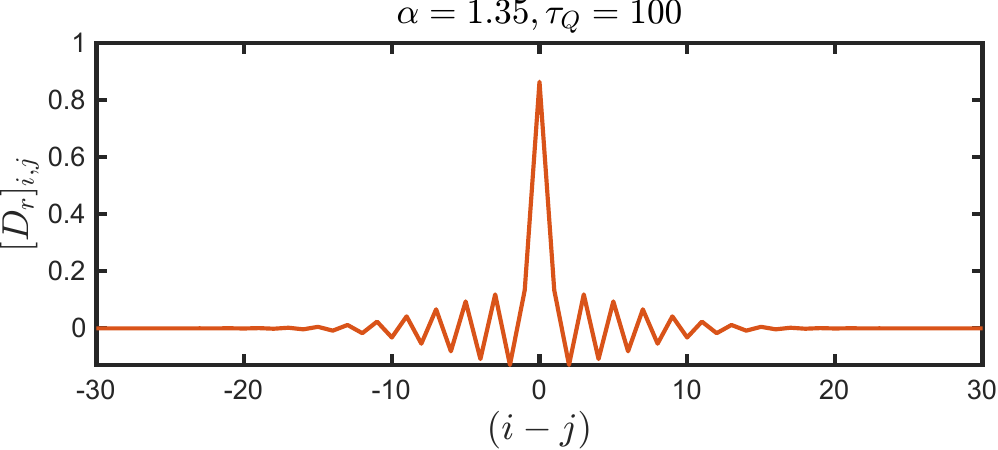}
    \caption{The matrix elements of $D_{r}$, as a function of separation are depicted in the plot, clearly illustrating their decay from the diagonal. While the plot is presented in the LR regime for the noiseless case, a similar trend is observed in the presence of noise.}
    \label{fig:sc_matrix_elements}
\end{figure}

 The longitudinal spin-spin correlation function $(\langle\sigma_{x}^{n}\sigma_{x}^{n+r}\rangle)$ can be expressed as a Majorana string operator and computed using Wick's theorem. This expression takes the form of a Toeplitz determinant, 
\begin{equation}
\begin{split}
       \mathcal{C}^{xx}(r) &=\langle B_{n}A_{n+1}B_{n+1}\cdot\cdot A_{n+r}\rangle\\
        &=D_{r}[g],
\end{split}
\end{equation}
where $D_{r}[g]$ represents an $r\times r$ Toeplitz determinant,
\begin{equation}
    \begin{split}
        D_{r}(g) &=\begin{vmatrix}
                 g(0) & g(-1) &\cdots\cdots\cdots & g(-r+1)\\
                 g(1) & g(0) &\cdots\cdots\cdots   & g(-r+2)\\
                 \vdots   &   \vdots   &\ddots    & \vdots  \\
                 g(r-1) & g(-r) &\cdots    & g(0)
                \end{vmatrix}
    \end{split}
\end{equation}
and $g(r)=\langle B_{n}A_{n+r} \rangle$ can be obtained via,
\begin{equation}
    \begin{split}
       g(r)
        &=\frac{1}{2\pi}\int_{-\pi}^{\pi}dk\, e^{ikr}\, (1-2p_{k}).
    \end{split}
\end{equation}
It is worth noting that while the two-point Majorana correlator shows crossover behavior, the longitudinal spin correlator decays exponentially at large separations for both the noisy and noiseless scenarios [see Fig.~(\ref{fig:spinxx})].
Since the matrix elements of $D_{r}(g)$ are suppressed strongly for entries far from the diagonal of the Toeplitz matrix [see Fig.~(\ref{fig:sc_matrix_elements})], the matrix resembles a banded matrix,  where only a small number of elements away from the diagonal  
 contribute significantly. Thus, resulting in exponential decay with SR correlation.

As the correlation length is defined as the inverse of the defect density, its behavior is closely linked to the dominant momentum modes contributing to defect formation~\cite{Levitov_2006,Zurek2007}. In the noiseless scenario, defect density is primarily governed by modes around $k=\pi$, leading to the KZ correlation length as $\xi_{0}\propto \sqrt{\tau_Q}$. This behavior is reflected in the rescaled correlation function shown in Fig.~\ref{fig:spinxx}(a). Moreover, as $\alpha$ is decreased, the defect density grows, signifying a reduction of the correlation length and a faster decay of correlations [see Fig.~\ref{fig:spinxx}(b)]. In contrast, under noisy conditions, defect formation is predominantly influenced by modes near the critical mode $k=0$. As the noise strength increases, the spin correlation function decays more rapidly, with numerical results indicating that the correlation length is 
 proportional to $(\eta_0^2 \tau_Q)^{-1}$ 
 (see the rescaled plot in Fig.~\ref{fig:spinxx}c). 
In this regime, increasing the range of LR interaction suppresses defect density, resulting in an extended correlation length and a slower decay of spin correlations, as illustrated in Fig.~\ref{fig:spinxx}(d). 

A similar behavior can be observed for emptiness probability, which belongs to the type-II category, as discussed in  Appendix C.

\section{FCS of kink distribution}
We will next consider Full counting statistics (FCS)  to analyze the probabilistic distribution of particle number fluctuations or other observables within a system. The approach  involves constructing a generating function or characteristic function, from which cumulants can be derived to quantify aspects like mean, variance, and higher-order correlations. 

Here we study the full counting statistics of topological defects (kink) for the  noiseless and noisy scenarios. While the mathematical formulation has been thoroughly discussed in previous studies ~\cite{delCampo2018,delcampo2021, MSingh_2023}, we provide here a brief overview for the sake of completeness. The probability distribution of any operator $\mathcal{N}$ is given by,
\begin{equation}
    P(n)=\langle \delta(\mathcal{N}-n) \rangle,
\end{equation}
and the distribution in terms of characteristic function can be written as,
\begin{equation}
    \begin{split}
        P(n)=\frac{1}{2\pi}\int_{-\pi}^{\pi}d\theta \bar{P}(\theta)e^{-in\theta},
    \end{split}
\end{equation}
and the characteristic function can be expressed as,
\begin{equation}
    \begin{split}
        \bar{P}(\theta) &=\text{Tr}[\rho e^{i\theta \mathcal{N}}].\\
    \end{split}
\end{equation}
The cumulants, $\kappa_s$, of $P(n)$ can be obtained via,
\begin{equation}
    \text{ln}[\bar{P}(\theta)]=\sum_{s=1}\frac{(i\theta)^s}{s!}\kappa_{s},
\end{equation}
where the cumulants are given by
\begin{equation}
    \begin{split}
        \kappa_{s} &=(-i)^s\frac{d^s}{d\theta^s}  ( \text{ln}[\bar{P}(\theta)])\Big|_{\theta=0}.\\
    \end{split}
\end{equation}
The number operator of topological defects $(\mathcal{N})$ (associated with the number of excited quasi-particles at the end of the quench protocol) is defined as
\begin{equation}
\begin{split}
 \mathcal{N} &=\sum_{k}\gamma^{\dagger}_{k}\gamma_{k},\\
\end{split}
 \end{equation}
where $\gamma_{k}$ is the quasi-particle operator diagonalizing the Hamiltonian with $\langle \gamma^{\dagger}_{k}\gamma_{k}\rangle=p_k$ and the characteristic function corresponding to the kink distribution,
\begin{equation}
    \begin{split}
        \bar{P}(\theta) &=\prod_{k}\text{Tr}[\rho_{k} e^{i\theta \gamma^{\dagger}_{k}\gamma_{k}}]\\
         &=\prod_{k}[1+(e^{i\theta}-1)p_{k}],\\
    \end{split}
\end{equation}
The logarithm of the characteristic function is given by
\begin{equation}
  \text{ln}(\bar{P}(\theta)) = \frac{1}{\pi}\int_{0}^{\pi}dk\,\text{ln}[1+(e^{i\theta}-1)p_{k}] 
\end{equation}
and the cumulants are expressed as
\begin{equation}
    \begin{split}
      \kappa_{1}  &= \frac{1}{\pi}\int_{0}^{\pi}dk\,   p_{k}\\
      \kappa_{2} &= \frac{1}{\pi}\int_{0}^{\pi}dk\,  p_{k}(1-   p_{k})\\
      \kappa_{3} &= \frac{1}{\pi}\int_{0}^{\pi}dk\, p_{k}(1-  p_{k})(1- 2 p_{k}) ,
    \end{split}
\end{equation}
where $\kappa_1$ represents the mean defect density, $\kappa_2$ quantifies the variance, indicating the spread of the distribution, and $\kappa_3$ measures the skewness, capturing the asymmetry in the kink distribution. 
\begin{figure}
    \centering
     \includegraphics[width=\columnwidth,height=4.5cm]{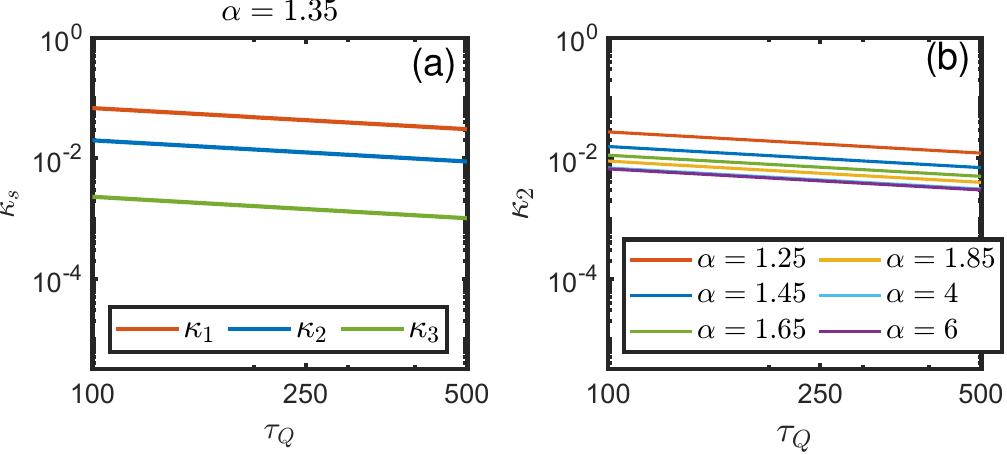}
    \includegraphics[width=\columnwidth,height=5cm]{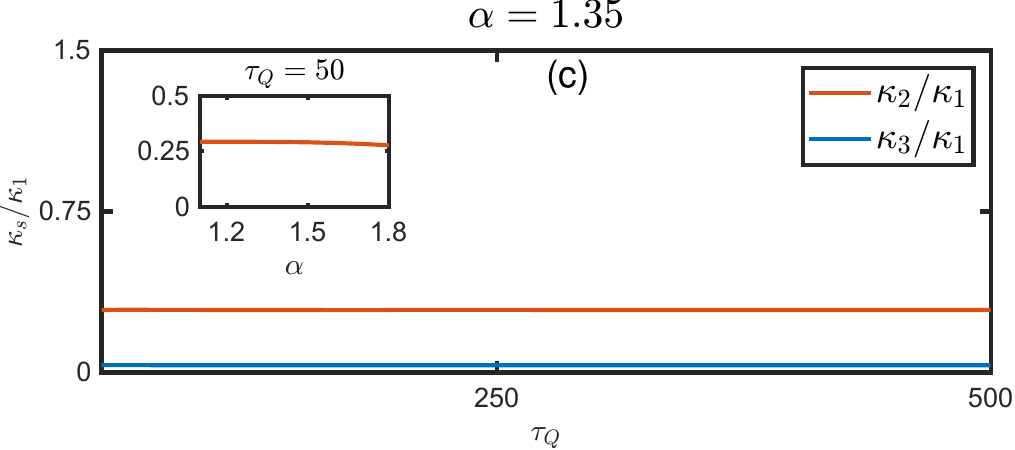}
    \caption{The influence of LR interactions on cumulants is illustrated in the figures.  Figure (a) presents the behavior of various cumulants as a function of the quench time $\tau_Q$ for a fixed value of $\alpha=1.35$. Figure (b) shows how the second cumulant varies with quench time for different $\alpha$ values, with similar trends observed for higher-order cumulants. Similar behavior can be observed for higher cumulants. Figure (c) presents the ratio of higher kink cumulants to the first cumulant as a function of quench time, while the inset shows this ratio as a function of $\alpha$ at a fixed quench time of $\tau_Q=50$.}
    \label{fig:noiseless_fcs}
\end{figure}
\begin{figure}
    \centering
    \includegraphics[width=\columnwidth,height=5cm]{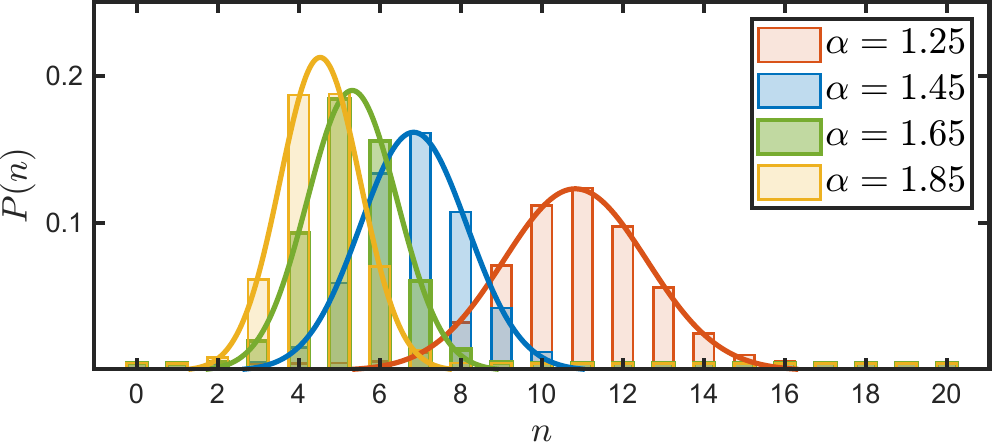}
    \caption{The plot presents the kink distribution for different $\alpha$ values at a fixed quench time of $\tau_Q=50$.}
    \label{fig:noiseless_kinkdistribution}
\end{figure}
In the noiseless scenario, all the cumulants follow the universal KZ scaling law, with higher-order cumulants being proportional to the first cumulant, $\kappa_{s\geq 2}\propto B(\alpha)\tau_{Q}^{-1/2}=\kappa_1$, which is evident from the Fig.~\ref{fig:noiseless_fcs}(a).  
The LR interaction influences the magnitude of the higher cumulants in a manner similar to the first cumulant  as shown for  $\kappa_2$ in Fig.~\ref{fig:noiseless_fcs}(b). The ratio of higher cumulants to the first cumulant remains invariant with respect to the quench time and LR interaction decaying parameter. Notably the higher cumulants $\kappa_s (s>2)$, are negligible compared to $\kappa_1$, as depicted in figure \ref{fig:noiseless_fcs}(c). This implies that kink distribution can be effectively characterized by $\kappa_1$ and $\kappa_2$ and can be approximated as the normal distribution (see the Fig.~\ref{fig:noiseless_kinkdistribution}). Furthermore, we observe that as the LR interaction increases, the second cumulant grows as shown in Fig.~\ref{fig:noiseless_fcs}(b), which results in the broadening of the distribution as can be seen from Fig.(\ref{fig:noiseless_kinkdistribution}).

\begin{figure}
    \centering
     \includegraphics[width=\columnwidth,height=5cm]{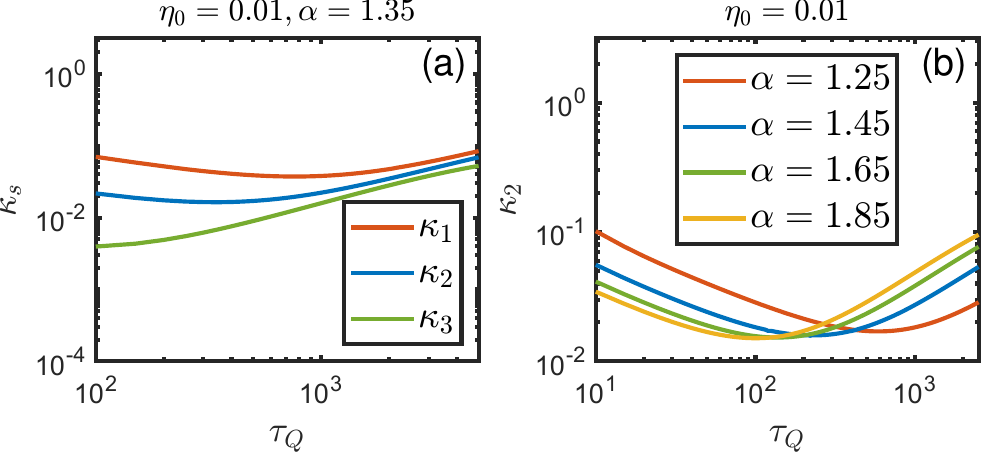}
    \caption{The behavior of the cumulants as a function of the quench time $\tau_Q$ have been shown for the noisy scenario, Figure.(a) illustrates the behavior of the cumulants for a fixed $\alpha=1.35$ and noise strength $\eta_0=0.01$. Figure.(b) depicts the behavior of the second cumulant $\kappa_2$ for different $\alpha$ values while keeping the noise strength fixed at $\eta_{0}=0.01$.}
    \label{fig:noise_fcs1}
\end{figure}
\begin{figure}
    \centering
     \includegraphics[width=\columnwidth,height=5cm]{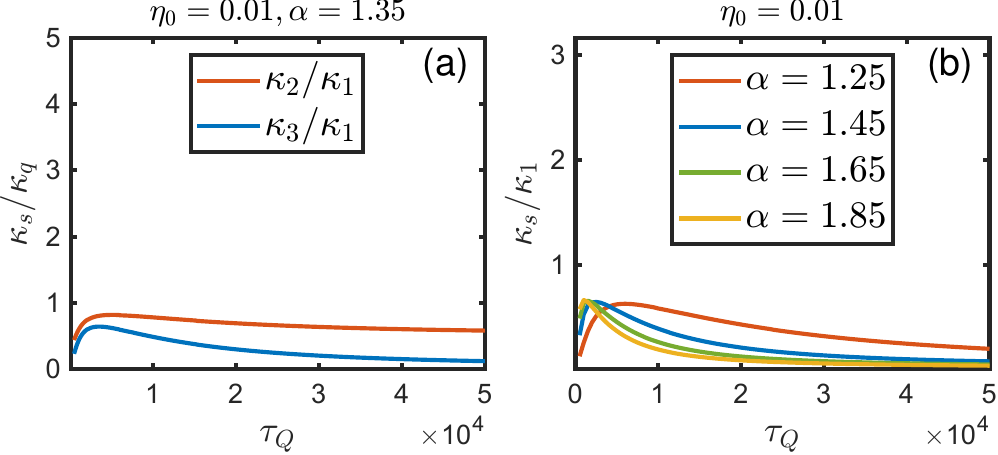}
    \caption{The behavior of the cumulants as a function of the quench time $\tau_Q$ have been shown for the noisy scenario, Figure.(a) illustrates the behavior of the cumulants for a fixed value of $\alpha=1.35$ and noise strength $\eta_{0}=0.01$. Figure.(b) depicts the behavior of the second cumulant $\kappa_{s\geq2}$ for different values of $\alpha$ while keeping the noise strength fixed at $\eta_{0}=0.01$.}
    \label{fig:noise_fcs2}
\end{figure}
\begin{figure}
    \centering
     \includegraphics[width=0.95\columnwidth,height=8cm]{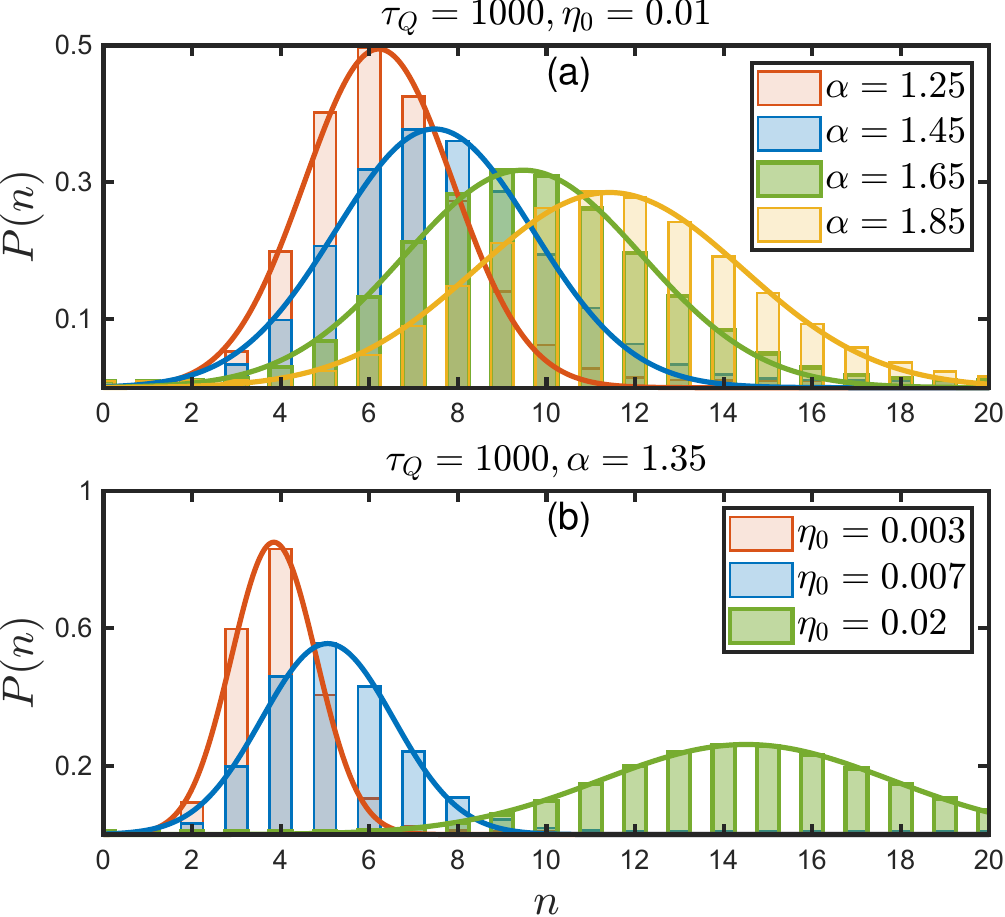}
    \caption{The kink distribution has been plotted for the noisy scenario, which shows the interplay {between noise and LR interaction decaying parameter}. Fig.(a) shows the kink distribution for different values of $\alpha$ with a fixed quench time $\tau_Q=100$ and noise strength $\eta_{0}=0.01$. In Fig.(b) the kink distribution is shown for different noise strengths for a fixed quench time $\tau_Q=1000$ and LR interaction strength at $\alpha=1.35$.}
    \label{fig:noisy_kinkdistribution}
\end{figure}

In the noisy scenario, all cumulants exhibit non-universal anti-Kibble-Zurek behavior as shown in Fig.~\ref{fig:noise_fcs1}(a). Notably, the second cumulant shows a crossover pattern similar to the mean defect density, where the crossover quench time increases with $\alpha$, which is depicted in Fig.~\ref{fig:noise_fcs1}(a). Specifically, beyond this crossover point, $\kappa_2$ decreases as the value of $\alpha$ decreases. The ratio $\kappa_2/\kappa_1$ remains constant with respect to the quench time in the large slow drive regime, while the ratio of higher-order cumulants  $(\kappa_{s>2})$ with $\kappa_1$ decays with increasing quench time. Moreover, this decay slows as the $\alpha$ increases (see Fig.~\ref{fig:noise_fcs2}). These findings indicate that, even in the presence of noise, the kink distribution is predominantly governed by $\kappa_1$ and $\kappa_2$ only, implying an approximately normal distribution (see Fig.~\ref{fig:noisy_kinkdistribution}). Moreover, as the value of $\alpha$ increases or the noise strength increases, the distribution broadens, reflecting enhanced defect fluctuations and deviations from the noiseless case, which is shown in the figure.~\ref{fig:noisy_kinkdistribution}(a)-(b).

\section{Discussion}
 Our study highlights the significant role of long-range interactions in shaping long-time dynamics under both noisy and noiseless conditions, revealing distinct contributions from different $k$ modes in each scenario.
 The impact of LR interaction is manifested in the defect density, its distribution, and spin correlation functions, resulting in distinct behavior between noisy and noiseless scenarios.
 In the noiseless scenario, our driving protocol in the long-range regime leads to dominant excitations near $k_c=\pi$, while modes around $k_c=0$ remain effectively adiabatic during slow drive. This behavior differs from the short-range case, where the modes around $k_c=0,\pi$ contribute symmetrically to the dynamics. 
The defect density follows the universal Kibble–Zurek scaling for $\alpha>1$, in the absence of noise. Notably, unlike recent studies~\cite{Anirban_dutta,N_defenu_TS}, we find that the scaling exponent remains insensitive to the LR decay exponent. In the presence of noise, the dominant mode in $p_k^{\eta_0}$, varies continuously with the decay exponent $\alpha$ in the range $1<\alpha<2$, whereas in the short-range limit, the maximum consistently occurs at $k=\pi/2$. Moreover the contribution of the dominant mode is suppressed as $\alpha$ decreases.
However, the defect density follows the AKZ scaling, similar to the short-range case, with the magnitude modified by the decay exponent $\alpha$. In the long-range regime, traversing the QCP at $k_c=0$ leads to a suppression of the two-point fermionic correlation function with increasing separation which is absent in the short-range case. From the analysis of the FCS of defects, we observe that defect distribution follows a Gaussian form in both noisy and noiseless scenarios with the variance exhibiting a clear dependence on the decay exponent, $\alpha$.  Our study extend recent studies on universality in TFIM-like models, shedding light on how LR interactions and noise influence higher-order cumulants and defect statistics.

In this work, we focus on the clean model however, exploring the disordered LR model and the influence of noise and disorder on defect production presents an exciting direction for future research. While our study emphasizes the fast-noise regime, where noise is introduced in the interaction term, investigating the effects of slow noise, in particular the role of noise correlation time needs further exploration. We also considered the case where noise is present in the transverse magnetic field and observed that the LZ probability exhibits qualitatively similar behavior to the transverse noise case.
Notably, such LR models have recently been explored in ion-trap experiments and digital quantum simulators under noisy conditions~\cite{Kandala2019, BWLi2023, DanielAzses2023, Gnezdilov2024, AntonioMandarino2024, Miessen2024}. Our driving protocol could potentially be implemented on such platforms, offering opportunities for further experimental validation.  
\appendix
\section{Expansion of $f_{\alpha}^{\infty}(k)$}
Expanding around $k=0$ and retaining only the leading-order terms, we obtain:

\begin{equation}
\begin{split}
f_{\alpha}^{\infty}(k) &\approx \cos{\frac{\alpha \pi}{2}} \frac{\Gamma(1-\alpha)}{\zeta(\alpha)} k^{\alpha-1} + \frac{\zeta(1-\alpha)}{\zeta(\alpha)} k, 
 \text{for } 1 < \alpha < 2,\\ 
 &\approx \frac{\zeta(1-\alpha)}{\zeta(\alpha)} k, 
\quad \text{for } \alpha > 2. 
\end{split} 
\end{equation}

Similarly, expanding around  $k=\pi$, we find:

\begin{equation} f_{\alpha}^{\infty}(k) \approx \frac{\zeta(1-\alpha)}{\zeta(\alpha)} (\pi - k), \quad \text{for any}\, \alpha > 1. \end{equation}

These expressions highlight the leading contributions to 
$f^{\infty}_{\alpha}(k)$ in the respective regions, capturing the key dependence on $k$ for different values of $\alpha$.

\section{Analytical derivation of survival probability}
From the Von-Neumann equation, we obtain
\begin{equation}
     \frac{d\rho_{k}^{\eta_{0}}}{dt} =-i[\mathcal{H}^{\eta_{0}}_{k},\rho_{k}^{\eta_{0}}]
\end{equation}
\begin{equation}
\label{eq.11}
    \begin{split}
        \dot{\tilde{\rho}}_{k}^{\eta_{0}}&=(\dot{\rho}_{k}^{\eta_{0}})_{11}-(\dot{\rho}_{k}^{\eta_{0}})_{22}\\
        &=2i(\rho_{k}^{\eta_{0}})_{12}(\mathcal{H}^{\eta_{0}}_k)_{21}-2i(\rho_{k}^{\eta_{0}})_{21}(\mathcal{H}^{\eta_{0}}_k)_{12}
    \end{split}
\end{equation}
\begin{equation}
\label{eq.12}
\begin{split}
    (\dot{\rho}_{k}^{\eta_{0}})_{12} &=-2iE_{k}(\rho_{k}^{\eta_{0}})_{12}+i(\mathcal{H}^{\eta_{0}}_{k})_{12}\tilde{\rho}^{\eta_{0}}_{k}\\
    (\rho_{k}^{\eta_{0}})_{12} &=i\int_{-\infty}^{t}\tilde{\rho}^{\eta_{0}}_{k}(\mathcal{H}^{\eta_{0}}_{k})_{12}e^{2i\int_{t_1}^{t}E_{k}(t_{2})dt_{2}}
\end{split}    
\end{equation}
where $(\mathcal{H}^{\eta_{0}}_k)_{12}=2i(f^{\infty}_{\alpha}+\eta_{}(t)f^{\infty}_{\alpha})$ and $(\mathcal{H}^{\eta_{0}}_k)_{21}=[(\mathcal{H}^{\eta_{0}}_{k})_{12}]^{*}$.\\
By substituting Eq.(\ref{eq.12}) in Eq.(\ref{eq.11}) we obtain,
\begin{multline}
     \dot{\tilde{\rho}}_{k}^{\eta_{0}}=-8(f_{\alpha}^{\infty})^2\int_{-\infty}^{t}dt_{1}e^{2i\int_{t_1}^{t}E_{k}(t_{2})dt_{2}}\tilde{\rho}^{\eta_{0}}_{k}(t_{1})\\
     -8(f_{\alpha}^{\infty})^2\int_{-\infty}^{t}dt_{1}e^{2i\int_{t_1}^{t}E_{k}(t_{2})dt_{2}}\eta(t)\eta(t_1)\tilde{\rho}^{\eta_{0}}_{k}(t_{1})+h.c.,
\end{multline}
where the cross term "$f_{\alpha}^{\infty}\eta(t)$" term is neglected since its noise average vanishes, $\langle\eta(t) \rangle=0$ and thus it does not contribute.\\
After taking the noise average $\langle\tilde{\rho}^{\eta_{0}}_{k}\rangle=\bar{\rho}_{k}$, we obtain
\begin{multline}
\label{noise_avr_diff_eqn}
     \dot{\bar{\rho}}_{k}=-8(f_{\alpha}^{\infty})^2\int^{t}_{-\infty}dt_{1}\cos{[2\beta(t-t_1)+(t^2-t^2_{1})\delta]}\bar{\rho}_{k}(t_{1})\\ -8(f_{\alpha}^{\infty})^2\int^{t}_{-\infty}dt_{1}\cos{[2\beta(t-t_1)+(t^2-t^2_{1})\delta]}\overline{\eta(t)\eta(t_1)}\bar{\rho}_{k}(t_{1})
\end{multline}
Considering fast noise approximation.
\begin{align}
    \cos[2\beta(t-t_{1})+\delta(t^2-t_{1}^2)] &\approx \cos[\bar{\omega}(t)(t-t_1)]
\end{align}
where $\bar{\omega}(t)=2(\beta+\delta t)$ with $\beta=-2g_{\alpha}^{\infty}$ and $\delta=1/\tau_Q$, and the differential Eq.(\ref{noise_avr_diff_eqn}) takes the form as follows
\begin{multline}
\label{final_diff_eqn}
       \dot{\bar{\rho}}_{k}=-8(f_{\alpha}^{\infty})^2\int^{t}_{-\infty}dt_{1}\cos{[2\beta(t-t_1)+(t^2-t^2_{1})\delta]}\bar{\rho}_{k}(t_{1})\\ -\Omega(t)\bar{\rho}_{k}(t_{}),
\end{multline}

where
\begin{align}
    \Omega[\bar{\omega}(t)]\equiv\Omega(t) &=4(f_{\alpha}^{\infty})^2(\Omega^{+}+\Omega^{-})
\end{align}
and  $\Omega^{\pm}[\bar{\omega}(t)]=\int_{-\infty}^{\infty}d\xi e^{\pm i\bar{\omega}\xi}R(|\xi|)=\frac{2\gamma\eta_{0}^2}{\gamma^2+\bar{\omega}(t)^2}$ are the power spectral density of noise capturing the environment effect with $R(|\xi|)=\eta_0^2e^{-\gamma|\xi|}$. By solving Eq.(\ref{final_diff_eqn})
\begin{equation}
\begin{split}
    \bar{\rho}_{k}(t)&=e^{-\int_{-\infty}^{t}\Omega(t')dt'}\rho_{k}^{0}(t)\\
\end{split} 
\end{equation}
where $\int_{-\infty}^{t}\Omega(t')dt'=\frac{8\pi(f_{\alpha}^{\infty})^2 R(0)}{\delta}[1+\frac{2}{\pi}\tan^{-1}(\frac{t\delta }{\gamma})]$. For $t\rightarrow \infty$, 
\begin{equation}
\begin{split}
     \int_{-\infty}^{\infty}\Omega(t')dt' &= \int_{-\infty}^{\infty} \frac{8\gamma\eta_{0}^2 (f^{\infty}_{\alpha})^2}{\gamma^2+\bar{\omega}(t')^2}dt'\\
     &=\frac{4\pi\eta_{0}^2 (f^{\infty}_{\alpha})^2}{\delta}
\end{split}
\end{equation}
the final density matrix $\rho_{k}$ is given by
\begin{equation}
   \bar{\rho}_{k}(\infty)=e^{-\frac{4\pi\eta_{0}^2(f^{\infty}_{\alpha})^2}{\delta}}(2e^{-2\pi\lambda}-1),
\end{equation}
where, $\lambda=\tau_{Q}(f^{\infty}_{\alpha})^2$. Hence, the survival probability of the system to be in ground state is
\begin{equation}
\begin{split}
     p^{\eta_{0}}_{k}&=\frac{1}{2}(1+e^{-4\pi\tau_{Q}\eta_{0}^2(f^{\infty}_{\alpha})^2}(2e^{-2\pi\tau_{Q}(f^{\infty}_{\alpha})^2}-1))\\
     & \approx \frac{1}{2}[1-e^{-4\pi\tau_{Q}\eta_{0}^2(f^{\infty}_{\alpha})^2}]+e^{-2\pi\tau_{Q}(f^{\infty}_{\alpha})^2}\\
     &= \bar{p}^{\eta}_{k}+p^{0}_{k}
\end{split}
\end{equation}

\section{Type-I and II spin correlation}
{\it Type-I spin correlation}\,:

For $r\gg 1$, the transverse spin-spin correlation functions can be expressed as,
\begin{equation}
\begin{split}
    \mathcal{C}^{zz}(r) &= \langle A_{n}B_{n}A_{n+r}B_{n+r}\rangle-\langle A_{n}B_{n}\rangle\langle A_{n+r}B_{n+r}   \rangle\\
    &\approx [G(r)]^2
\end{split}
\end{equation}
and the kink correlator, $\rho^{zz}(r)$ can be written as follows,
\begin{equation}
    \begin{split}
        \rho^{zz}(r) &=\langle \sigma_{z}^{n}\sigma_{z}^{n+1}\sigma_{z}^{n+r}\sigma_{z}^{n+r+1}\rangle -\langle  \sigma_{z}^{n}\sigma_{z}^{n+1}\rangle^{2}\\
        &=\Delta_{1}(r)-\Delta_{2}^{2}
    \end{split}
\end{equation}
where  $\Delta_{1}(r)$ is defined as
\begin{equation}
    \begin{split}
        \Delta_{1}(r) &=\langle A_{n}B_{n}A_{n+1}B_{n+1}A_{n+r}B_{n+r}A_{n+r+1}B_{n+r+1} \rangle\\
       &= \begin{vmatrix}
                 G(0) & G(1) & G(r) & G(r+1)\\
                 G(-1) & G(0) & G(r-1) & G(r)\\
                 G(-r) & G(-r+1)   & G(0) & G(1) \\
                 G(-r-1) & G(-r)   & G(-1)  &G(0)
                \end{vmatrix},\\
    \end{split}
\end{equation}
and $\Delta_{2}$ is given by
\begin{equation}
    \begin{split}
        \Delta_{2}&=\langle A_{n}B_{n}A_{n+1}B_{n+1}\rangle\\ 
            &=\begin{vmatrix}
                 G(0) & G(1) \\
                 G(-1) & G(0)\\
                \end{vmatrix}.
    \end{split}
\end{equation}
where for slow drive scenario $G(r\pm1)\approx G(r)$, and therefore we obtain
\begin{equation} 
        \rho^{zz}(r) \approx -[G(0)]^2[G(r)]^2
        \label{kinkcorr}
\end{equation}
where, the negative sign in Eq.\ref{kinkcorr}, signifying the antibunching effect.

{\it Type-II spin correlation}\,:

 The string correlator, $ \mathcal{E}(r)$,
 representing the emptiness probability can also be derived using Wick's theorem and can be written as~\cite{Abanov_Franchini_2005}
\begin{equation}
    \begin{split}
        \mathcal{E}(r) &=\langle \prod_{l=1}^{r}\frac{(1-\sigma_{z}^{l})}{2} \rangle\\
        &=\langle\prod_{l} c^{\dagger}_{l}c_{l}   \rangle\\
        &=|\text{det} S|
    \end{split}
\end{equation}
where the elements of $S$ are given by,
\begin{equation}
    \begin{split}
                S_{mn}&=\langle c^{\dagger}_{m} c_{n} \rangle\\
                &=\frac{1}{2\pi}\int_{-\pi}^{\pi}dk\, e^{i(m-n)k}(1-p_{k}),
    \end{split}
\end{equation}
and, $\langle c^{\dagger}_{n}c^{\dagger}_{m} \rangle, \langle c_{n}c_{m} \rangle=0$.

\bibliography{LR_model_KZ}

\begin{thebibliography}{57}%
\makeatletter
\providecommand \@ifxundefined [1]{%
 \@ifx{#1\undefined}
}%
\providecommand \@ifnum [1]{%
 \ifnum #1\expandafter \@firstoftwo
 \else \expandafter \@secondoftwo
 \fi
}%
\providecommand \@ifx [1]{%
 \ifx #1\expandafter \@firstoftwo
 \else \expandafter \@secondoftwo
 \fi
}%
\providecommand \natexlab [1]{#1}%
\providecommand \enquote  [1]{``#1''}%
\providecommand \bibnamefont  [1]{#1}%
\providecommand \bibfnamefont [1]{#1}%
\providecommand \citenamefont [1]{#1}%
\providecommand \href@noop [0]{\@secondoftwo}%
\providecommand \href [0]{\begingroup \@sanitize@url \@href}%
\providecommand \@href[1]{\@@startlink{#1}\@@href}%
\providecommand \@@href[1]{\endgroup#1\@@endlink}%
\providecommand \@sanitize@url [0]{\catcode `\\12\catcode `\$12\catcode
  `\&12\catcode `\#12\catcode `\^12\catcode `\_12\catcode `\%12\relax}%
\providecommand \@@startlink[1]{}%
\providecommand \@@endlink[0]{}%
\providecommand \url  [0]{\begingroup\@sanitize@url \@url }%
\providecommand \@url [1]{\endgroup\@href {#1}{\urlprefix }}%
\providecommand \urlprefix  [0]{URL }%
\providecommand \Eprint [0]{\href }%
\providecommand \doibase [0]{https://doi.org/}%
\providecommand \selectlanguage [0]{\@gobble}%
\providecommand \bibinfo  [0]{\@secondoftwo}%
\providecommand \bibfield  [0]{\@secondoftwo}%
\providecommand \translation [1]{[#1]}%
\providecommand \BibitemOpen [0]{}%
\providecommand \bibitemStop [0]{}%
\providecommand \bibitemNoStop [0]{.\EOS\space}%
\providecommand \EOS [0]{\spacefactor3000\relax}%
\providecommand \BibitemShut  [1]{\csname bibitem#1\endcsname}%
\let\auto@bib@innerbib\@empty
\bibitem [{\citenamefont {Campa}\ \emph {et~al.}(2009)\citenamefont {Campa},
  \citenamefont {Dauxois},\ and\ \citenamefont {Ruffo}}]{Campa2009}%
  \BibitemOpen
  \bibfield  {author} {\bibinfo {author} {\bibfnamefont {A.}~\bibnamefont
  {Campa}}, \bibinfo {author} {\bibfnamefont {T.}~\bibnamefont {Dauxois}},\
  and\ \bibinfo {author} {\bibfnamefont {S.}~\bibnamefont {Ruffo}},\ }\bibfield
   {title} {\bibinfo {title} {Statistical mechanics and dynamics of solvable
  models with long-range interactions},\ }\bibfield  {journal} {\bibinfo
  {journal} {Physics Reports}\ }\href {https://doi.org/physrep.2009.07.001}
  {physrep.2009.07.001} (\bibinfo {year} {2009})\BibitemShut {NoStop}%
\bibitem [{\citenamefont {Defenu}\ \emph {et~al.}(2023)\citenamefont {Defenu},
  \citenamefont {Donner}, \citenamefont {Macr\`{\i}}, \citenamefont {Pagano},
  \citenamefont {Ruffo},\ and\ \citenamefont {Trombettoni}}]{NicoloDefenu2023}%
  \BibitemOpen
  \bibfield  {author} {\bibinfo {author} {\bibfnamefont {N.}~\bibnamefont
  {Defenu}}, \bibinfo {author} {\bibfnamefont {T.}~\bibnamefont {Donner}},
  \bibinfo {author} {\bibfnamefont {T.}~\bibnamefont {Macr\`{\i}}}, \bibinfo
  {author} {\bibfnamefont {G.}~\bibnamefont {Pagano}}, \bibinfo {author}
  {\bibfnamefont {S.}~\bibnamefont {Ruffo}},\ and\ \bibinfo {author}
  {\bibfnamefont {A.}~\bibnamefont {Trombettoni}},\ }\bibfield  {title}
  {\bibinfo {title} {Long-range interacting quantum systems},\ }\href
  {https://doi.org/10.1103/RevModPhys.95.035002} {\bibfield  {journal}
  {\bibinfo  {journal} {Rev. Mod. Phys.}\ }\textbf {\bibinfo {volume} {95}},\
  \bibinfo {pages} {035002} (\bibinfo {year} {2023})}\BibitemShut {NoStop}%
\bibitem [{\citenamefont {Blatt}\ and\ \citenamefont {Roos}(2012)}]{Blatt2012}%
  \BibitemOpen
  \bibfield  {author} {\bibinfo {author} {\bibfnamefont {R.}~\bibnamefont
  {Blatt}}\ and\ \bibinfo {author} {\bibfnamefont {C.~F.}\ \bibnamefont
  {Roos}},\ }\bibfield  {title} {\bibinfo {title} {Quantum simulations with
  trapped ions},\ }\href {https://doi.org/10.1038/nphys2252} {\bibfield
  {journal} {\bibinfo  {journal} {Nature Physics}\ }\textbf {\bibinfo {volume}
  {8}},\ \bibinfo {pages} {277} (\bibinfo {year} {2012})}\BibitemShut {NoStop}%
\bibitem [{\citenamefont {Votto}\ \emph {et~al.}(2024)\citenamefont {Votto},
  \citenamefont {Zeiher},\ and\ \citenamefont
  {Vermersch}}]{Votto2024universalquantum}%
  \BibitemOpen
  \bibfield  {author} {\bibinfo {author} {\bibfnamefont {M.}~\bibnamefont
  {Votto}}, \bibinfo {author} {\bibfnamefont {J.}~\bibnamefont {Zeiher}},\ and\
  \bibinfo {author} {\bibfnamefont {B.}~\bibnamefont {Vermersch}},\ }\bibfield
  {title} {\bibinfo {title} {Universal quantum processors in spin systems via
  robust local pulse sequences},\ }\href
  {https://doi.org/10.22331/q-2024-10-29-1513} {\bibfield  {journal} {\bibinfo
  {journal} {{Quantum}}\ }\textbf {\bibinfo {volume} {8}},\ \bibinfo {pages}
  {1513} (\bibinfo {year} {2024})}\BibitemShut {NoStop}%
\bibitem [{\citenamefont {Browaeys}\ and\ \citenamefont
  {Lahaye}(2020)}]{Browaeys2020}%
  \BibitemOpen
  \bibfield  {author} {\bibinfo {author} {\bibfnamefont {A.}~\bibnamefont
  {Browaeys}}\ and\ \bibinfo {author} {\bibfnamefont {T.}~\bibnamefont
  {Lahaye}},\ }\bibfield  {title} {\bibinfo {title} {Many-body physics with
  individually controlled rydberg atoms},\ }\href
  {https://doi.org/10.1038/s41567-019-0733-z} {\bibfield  {journal} {\bibinfo
  {journal} {Nature Physics}\ }\textbf {\bibinfo {volume} {16}},\ \bibinfo
  {pages} {132} (\bibinfo {year} {2020})}\BibitemShut {NoStop}%
\bibitem [{\citenamefont {Monroe}\ \emph {et~al.}(2021)\citenamefont {Monroe},
  \citenamefont {Campbell}, \citenamefont {Duan}, \citenamefont {Gong},
  \citenamefont {Gorshkov}, \citenamefont {Hess}, \citenamefont {Islam},
  \citenamefont {Kim}, \citenamefont {Linke}, \citenamefont {Pagano},
  \citenamefont {Richerme}, \citenamefont {Senko},\ and\ \citenamefont
  {Yao}}]{Yao2021}%
  \BibitemOpen
  \bibfield  {author} {\bibinfo {author} {\bibfnamefont {C.}~\bibnamefont
  {Monroe}}, \bibinfo {author} {\bibfnamefont {W.~C.}\ \bibnamefont
  {Campbell}}, \bibinfo {author} {\bibfnamefont {L.-M.}\ \bibnamefont {Duan}},
  \bibinfo {author} {\bibfnamefont {Z.-X.}\ \bibnamefont {Gong}}, \bibinfo
  {author} {\bibfnamefont {A.~V.}\ \bibnamefont {Gorshkov}}, \bibinfo {author}
  {\bibfnamefont {P.~W.}\ \bibnamefont {Hess}}, \bibinfo {author}
  {\bibfnamefont {R.}~\bibnamefont {Islam}}, \bibinfo {author} {\bibfnamefont
  {K.}~\bibnamefont {Kim}}, \bibinfo {author} {\bibfnamefont {N.~M.}\
  \bibnamefont {Linke}}, \bibinfo {author} {\bibfnamefont {G.}~\bibnamefont
  {Pagano}}, \bibinfo {author} {\bibfnamefont {P.}~\bibnamefont {Richerme}},
  \bibinfo {author} {\bibfnamefont {C.}~\bibnamefont {Senko}},\ and\ \bibinfo
  {author} {\bibfnamefont {N.~Y.}\ \bibnamefont {Yao}},\ }\bibfield  {title}
  {\bibinfo {title} {Programmable quantum simulations of spin systems with
  trapped ions},\ }\href {https://doi.org/10.1103/RevModPhys.93.025001}
  {\bibfield  {journal} {\bibinfo  {journal} {Rev. Mod. Phys.}\ }\textbf
  {\bibinfo {volume} {93}},\ \bibinfo {pages} {025001} (\bibinfo {year}
  {2021})}\BibitemShut {NoStop}%
\bibitem [{\citenamefont {M\"unstermann}\ \emph {et~al.}(2000)\citenamefont
  {M\"unstermann}, \citenamefont {Fischer}, \citenamefont {Maunz},
  \citenamefont {Pinkse},\ and\ \citenamefont {Rempe}}]{Rempe2000}%
  \BibitemOpen
  \bibfield  {author} {\bibinfo {author} {\bibfnamefont {P.}~\bibnamefont
  {M\"unstermann}}, \bibinfo {author} {\bibfnamefont {T.}~\bibnamefont
  {Fischer}}, \bibinfo {author} {\bibfnamefont {P.}~\bibnamefont {Maunz}},
  \bibinfo {author} {\bibfnamefont {P.~W.~H.}\ \bibnamefont {Pinkse}},\ and\
  \bibinfo {author} {\bibfnamefont {G.}~\bibnamefont {Rempe}},\ }\bibfield
  {title} {\bibinfo {title} {Observation of cavity-mediated long-range light
  forces between strongly coupled atoms},\ }\href
  {https://doi.org/10.1103/PhysRevLett.84.4068} {\bibfield  {journal} {\bibinfo
   {journal} {Phys. Rev. Lett.}\ }\textbf {\bibinfo {volume} {84}},\ \bibinfo
  {pages} {4068} (\bibinfo {year} {2000})}\BibitemShut {NoStop}%
\bibitem [{\citenamefont {Saffman}\ \emph {et~al.}(2010)\citenamefont
  {Saffman}, \citenamefont {Walker},\ and\ \citenamefont
  {M\o{}lmer}}]{Saffman2010}%
  \BibitemOpen
  \bibfield  {author} {\bibinfo {author} {\bibfnamefont {M.}~\bibnamefont
  {Saffman}}, \bibinfo {author} {\bibfnamefont {T.~G.}\ \bibnamefont
  {Walker}},\ and\ \bibinfo {author} {\bibfnamefont {K.}~\bibnamefont
  {M\o{}lmer}},\ }\bibfield  {title} {\bibinfo {title} {Quantum information
  with rydberg atoms},\ }\href {https://doi.org/10.1103/RevModPhys.82.2313}
  {\bibfield  {journal} {\bibinfo  {journal} {Rev. Mod. Phys.}\ }\textbf
  {\bibinfo {volume} {82}},\ \bibinfo {pages} {2313} (\bibinfo {year}
  {2010})}\BibitemShut {NoStop}%
\bibitem [{\citenamefont {Schachenmayer}\ \emph {et~al.}(2013)\citenamefont
  {Schachenmayer}, \citenamefont {Lanyon}, \citenamefont {Roos},\ and\
  \citenamefont {Daley}}]{Schachenmayer2013}%
  \BibitemOpen
  \bibfield  {author} {\bibinfo {author} {\bibfnamefont {J.}~\bibnamefont
  {Schachenmayer}}, \bibinfo {author} {\bibfnamefont {B.~P.}\ \bibnamefont
  {Lanyon}}, \bibinfo {author} {\bibfnamefont {C.~F.}\ \bibnamefont {Roos}},\
  and\ \bibinfo {author} {\bibfnamefont {A.~J.}\ \bibnamefont {Daley}},\
  }\bibfield  {title} {\bibinfo {title} {Entanglement growth in quench dynamics
  with variable range interactions},\ }\href
  {https://doi.org/10.1103/PhysRevX.3.031015} {\bibfield  {journal} {\bibinfo
  {journal} {Phys. Rev. X}\ }\textbf {\bibinfo {volume} {3}},\ \bibinfo {pages}
  {031015} (\bibinfo {year} {2013})}\BibitemShut {NoStop}%
\bibitem [{\citenamefont {Jaschke}\ \emph {et~al.}(2017)\citenamefont
  {Jaschke}, \citenamefont {Maeda}, \citenamefont {Whalen}, \citenamefont
  {Wall},\ and\ \citenamefont {Carr}}]{Jaschke_2017}%
  \BibitemOpen
  \bibfield  {author} {\bibinfo {author} {\bibfnamefont {D.}~\bibnamefont
  {Jaschke}}, \bibinfo {author} {\bibfnamefont {K.}~\bibnamefont {Maeda}},
  \bibinfo {author} {\bibfnamefont {J.~D.}\ \bibnamefont {Whalen}}, \bibinfo
  {author} {\bibfnamefont {M.~L.}\ \bibnamefont {Wall}},\ and\ \bibinfo
  {author} {\bibfnamefont {L.~D.}\ \bibnamefont {Carr}},\ }\bibfield  {title}
  {\bibinfo {title} {Critical phenomena and kibble–zurek scaling in the
  long-range quantum ising chain},\ }\href
  {https://doi.org/10.1088/1367-2630/aa65bc} {\bibfield  {journal} {\bibinfo
  {journal} {New Journal of Physics}\ }\textbf {\bibinfo {volume} {19}},\
  \bibinfo {pages} {033032} (\bibinfo {year} {2017})}\BibitemShut {NoStop}%
\bibitem [{\citenamefont {Igl\'oi}\ \emph {et~al.}(2018)\citenamefont
  {Igl\'oi}, \citenamefont {Bla\ss{}}, \citenamefont {Ro\'osz},\ and\
  \citenamefont {Rieger}}]{FerencIgloi2018}%
  \BibitemOpen
  \bibfield  {author} {\bibinfo {author} {\bibfnamefont {F.}~\bibnamefont
  {Igl\'oi}}, \bibinfo {author} {\bibfnamefont {B.}~\bibnamefont {Bla\ss{}}},
  \bibinfo {author} {\bibfnamefont {G.~m.~H.}\ \bibnamefont {Ro\'osz}},\ and\
  \bibinfo {author} {\bibfnamefont {H.}~\bibnamefont {Rieger}},\ }\bibfield
  {title} {\bibinfo {title} {Quantum xx model with competing short- and
  long-range interactions: Phases and phase transitions in and out of
  equilibrium},\ }\href {https://doi.org/10.1103/PhysRevB.98.184415} {\bibfield
   {journal} {\bibinfo  {journal} {Phys. Rev. B}\ }\textbf {\bibinfo {volume}
  {98}},\ \bibinfo {pages} {184415} (\bibinfo {year} {2018})}\BibitemShut
  {NoStop}%
\bibitem [{\citenamefont {Uhrich}\ \emph {et~al.}(2020)\citenamefont {Uhrich},
  \citenamefont {Defenu}, \citenamefont {Jafari},\ and\ \citenamefont
  {Halimeh}}]{Uhrich2020}%
  \BibitemOpen
  \bibfield  {author} {\bibinfo {author} {\bibfnamefont {P.}~\bibnamefont
  {Uhrich}}, \bibinfo {author} {\bibfnamefont {N.}~\bibnamefont {Defenu}},
  \bibinfo {author} {\bibfnamefont {R.}~\bibnamefont {Jafari}},\ and\ \bibinfo
  {author} {\bibfnamefont {J.~C.}\ \bibnamefont {Halimeh}},\ }\bibfield
  {title} {\bibinfo {title} {Out-of-equilibrium phase diagram of long-range
  superconductors},\ }\href {https://doi.org/10.1103/PhysRevB.101.245148}
  {\bibfield  {journal} {\bibinfo  {journal} {Phys. Rev. B}\ }\textbf {\bibinfo
  {volume} {101}},\ \bibinfo {pages} {245148} (\bibinfo {year}
  {2020})}\BibitemShut {NoStop}%
\bibitem [{\citenamefont {Yang}\ \emph {et~al.}(2022)\citenamefont {Yang},
  \citenamefont {Pang}, \citenamefont {del Campo},\ and\ \citenamefont
  {Jordan}}]{delCampo2022}%
  \BibitemOpen
  \bibfield  {author} {\bibinfo {author} {\bibfnamefont {J.}~\bibnamefont
  {Yang}}, \bibinfo {author} {\bibfnamefont {S.}~\bibnamefont {Pang}}, \bibinfo
  {author} {\bibfnamefont {A.}~\bibnamefont {del Campo}},\ and\ \bibinfo
  {author} {\bibfnamefont {A.~N.}\ \bibnamefont {Jordan}},\ }\bibfield  {title}
  {\bibinfo {title} {Super-heisenberg scaling in hamiltonian parameter
  estimation in the long-range kitaev chain},\ }\href
  {https://doi.org/10.1103/PhysRevResearch.4.013133} {\bibfield  {journal}
  {\bibinfo  {journal} {Phys. Rev. Res.}\ }\textbf {\bibinfo {volume} {4}},\
  \bibinfo {pages} {013133} (\bibinfo {year} {2022})}\BibitemShut {NoStop}%
\bibitem [{\citenamefont {Zurek}(1985)}]{Zurek1985}%
  \BibitemOpen
  \bibfield  {author} {\bibinfo {author} {\bibfnamefont {W.~H.}\ \bibnamefont
  {Zurek}},\ }\bibfield  {title} {\bibinfo {title} {Cosmological experiments in
  superfluid helium?},\ }\href {https://doi.org/10.1038/317505a0} {\bibfield
  {journal} {\bibinfo  {journal} {Nature}\ }\textbf {\bibinfo {volume} {317}},\
  \bibinfo {pages} {505} (\bibinfo {year} {1985})}\BibitemShut {NoStop}%
\bibitem [{\citenamefont {Zurek}(1996)}]{ZUREK1996}%
  \BibitemOpen
  \bibfield  {author} {\bibinfo {author} {\bibfnamefont {W.}~\bibnamefont
  {Zurek}},\ }\bibfield  {title} {\bibinfo {title} {Cosmological experiments in
  condensed matter systems},\ }\href
  {https://doi.org/https://doi.org/10.1016/S0370-1573(96)00009-9} {\bibfield
  {journal} {\bibinfo  {journal} {Physics Reports}\ }\textbf {\bibinfo {volume}
  {276}},\ \bibinfo {pages} {177} (\bibinfo {year} {1996})}\BibitemShut
  {NoStop}%
\bibitem [{\citenamefont {Dziarmaga}(2010)}]{Dziarmaga2010}%
  \BibitemOpen
  \bibfield  {author} {\bibinfo {author} {\bibfnamefont {J.}~\bibnamefont
  {Dziarmaga}},\ }\bibfield  {title} {\bibinfo {title} {Dynamics of a quantum
  phase transition and relaxation to a steady state},\ }\href
  {https://doi.org/10.1080/00018732.2010.514702} {\bibfield  {journal}
  {\bibinfo  {journal} {Advances in Physics}\ }\textbf {\bibinfo {volume}
  {59}},\ \bibinfo {pages} {1063} (\bibinfo {year} {2010})}\BibitemShut
  {NoStop}%
\bibitem [{\citenamefont {Kolodrubetz}\ \emph {et~al.}(2012)\citenamefont
  {Kolodrubetz}, \citenamefont {Clark},\ and\ \citenamefont
  {Huse}}]{Kolodrubetz2012}%
  \BibitemOpen
  \bibfield  {author} {\bibinfo {author} {\bibfnamefont {M.}~\bibnamefont
  {Kolodrubetz}}, \bibinfo {author} {\bibfnamefont {B.~K.}\ \bibnamefont
  {Clark}},\ and\ \bibinfo {author} {\bibfnamefont {D.~A.}\ \bibnamefont
  {Huse}},\ }\bibfield  {title} {\bibinfo {title} {Nonequilibrium dynamic
  critical scaling of the quantum ising chain},\ }\href
  {https://doi.org/10.1103/PhysRevLett.109.015701} {\bibfield  {journal}
  {\bibinfo  {journal} {Phys. Rev. Lett.}\ }\textbf {\bibinfo {volume} {109}},\
  \bibinfo {pages} {015701} (\bibinfo {year} {2012})}\BibitemShut {NoStop}%
\bibitem [{\citenamefont {del Campo}\ and\ \citenamefont
  {Zurek}(2014)}]{delcampo2014}%
  \BibitemOpen
  \bibfield  {author} {\bibinfo {author} {\bibfnamefont {A.}~\bibnamefont {del
  Campo}}\ and\ \bibinfo {author} {\bibfnamefont {W.~H.}\ \bibnamefont
  {Zurek}},\ }\bibfield  {title} {\bibinfo {title} {Universality of phase
  transition dynamics: Topological defects from symmetry breaking},\ }\href
  {https://doi.org/10.1142/S0217751X1430018X} {\bibfield  {journal} {\bibinfo
  {journal} {International Journal of Modern Physics A}\ }\textbf {\bibinfo
  {volume} {29}},\ \bibinfo {pages} {1430018} (\bibinfo {year}
  {2014})}\BibitemShut {NoStop}%
\bibitem [{\citenamefont {Puebla}\ \emph {et~al.}(2020)\citenamefont {Puebla},
  \citenamefont {Smirne}, \citenamefont {Huelga},\ and\ \citenamefont
  {Plenio}}]{RicardoPuebla2020}%
  \BibitemOpen
  \bibfield  {author} {\bibinfo {author} {\bibfnamefont {R.}~\bibnamefont
  {Puebla}}, \bibinfo {author} {\bibfnamefont {A.}~\bibnamefont {Smirne}},
  \bibinfo {author} {\bibfnamefont {S.~F.}\ \bibnamefont {Huelga}},\ and\
  \bibinfo {author} {\bibfnamefont {M.~B.}\ \bibnamefont {Plenio}},\ }\bibfield
   {title} {\bibinfo {title} {Universal anti-kibble-zurek scaling in fully
  connected systems},\ }\href {https://doi.org/10.1103/PhysRevLett.124.230602}
  {\bibfield  {journal} {\bibinfo  {journal} {Phys. Rev. Lett.}\ }\textbf
  {\bibinfo {volume} {124}},\ \bibinfo {pages} {230602} (\bibinfo {year}
  {2020})}\BibitemShut {NoStop}%
\bibitem [{\citenamefont {Roberts}\ and\ \citenamefont
  {Clerk}(2023)}]{AClerk2023}%
  \BibitemOpen
  \bibfield  {author} {\bibinfo {author} {\bibfnamefont {D.}~\bibnamefont
  {Roberts}}\ and\ \bibinfo {author} {\bibfnamefont {A.~A.}\ \bibnamefont
  {Clerk}},\ }\bibfield  {title} {\bibinfo {title} {Exact solution of the
  infinite-range dissipative transverse-field ising model},\ }\href
  {https://doi.org/10.1103/PhysRevLett.131.190403} {\bibfield  {journal}
  {\bibinfo  {journal} {Phys. Rev. Lett.}\ }\textbf {\bibinfo {volume} {131}},\
  \bibinfo {pages} {190403} (\bibinfo {year} {2023})}\BibitemShut {NoStop}%
\bibitem [{\citenamefont {King}\ \emph {et~al.}(2023)\citenamefont {King},
  \citenamefont {Kriel},\ and\ \citenamefont {Kastner}}]{Kastner2023}%
  \BibitemOpen
  \bibfield  {author} {\bibinfo {author} {\bibfnamefont {E.~C.}\ \bibnamefont
  {King}}, \bibinfo {author} {\bibfnamefont {J.~N.}\ \bibnamefont {Kriel}},\
  and\ \bibinfo {author} {\bibfnamefont {M.}~\bibnamefont {Kastner}},\
  }\bibfield  {title} {\bibinfo {title} {Universal cooling dynamics toward a
  quantum critical point},\ }\href
  {https://doi.org/10.1103/PhysRevLett.130.050401} {\bibfield  {journal}
  {\bibinfo  {journal} {Phys. Rev. Lett.}\ }\textbf {\bibinfo {volume} {130}},\
  \bibinfo {pages} {050401} (\bibinfo {year} {2023})}\BibitemShut {NoStop}%
\bibitem [{\citenamefont {Mattes}\ \emph {et~al.}(2024)\citenamefont {Mattes},
  \citenamefont {Lesanovsky},\ and\ \citenamefont
  {Carollo}}]{FedericoCarollo2024}%
  \BibitemOpen
  \bibfield  {author} {\bibinfo {author} {\bibfnamefont {R.}~\bibnamefont
  {Mattes}}, \bibinfo {author} {\bibfnamefont {I.}~\bibnamefont {Lesanovsky}},\
  and\ \bibinfo {author} {\bibfnamefont {F.}~\bibnamefont {Carollo}},\ }\href
  {https://arxiv.org/abs/2407.02141} {\bibinfo {title} {Long-range interacting
  systems are locally non-interacting}} (\bibinfo {year} {2024}),\ \Eprint
  {https://arxiv.org/abs/2407.02141} {arXiv:2407.02141 [cond-mat.stat-mech]}
  \BibitemShut {NoStop}%
\bibitem [{\citenamefont {Baghran}\ \emph {et~al.}(2024)\citenamefont
  {Baghran}, \citenamefont {Jafari},\ and\ \citenamefont
  {Langari}}]{Langari2024}%
  \BibitemOpen
  \bibfield  {author} {\bibinfo {author} {\bibfnamefont {R.}~\bibnamefont
  {Baghran}}, \bibinfo {author} {\bibfnamefont {R.}~\bibnamefont {Jafari}},\
  and\ \bibinfo {author} {\bibfnamefont {A.}~\bibnamefont {Langari}},\
  }\bibfield  {title} {\bibinfo {title} {Competition of long-range interactions
  and noise at a ramped quench dynamical quantum phase transition: The case of
  the long-range pairing kitaev chain},\ }\href
  {https://doi.org/10.1103/PhysRevB.110.064302} {\bibfield  {journal} {\bibinfo
   {journal} {Phys. Rev. B}\ }\textbf {\bibinfo {volume} {110}},\ \bibinfo
  {pages} {064302} (\bibinfo {year} {2024})}\BibitemShut {NoStop}%
\bibitem [{\citenamefont {Caneva}\ \emph {et~al.}(2008)\citenamefont {Caneva},
  \citenamefont {Fazio},\ and\ \citenamefont {Santoro}}]{RosarioFazio2008}%
  \BibitemOpen
  \bibfield  {author} {\bibinfo {author} {\bibfnamefont {T.}~\bibnamefont
  {Caneva}}, \bibinfo {author} {\bibfnamefont {R.}~\bibnamefont {Fazio}},\ and\
  \bibinfo {author} {\bibfnamefont {G.~E.}\ \bibnamefont {Santoro}},\
  }\bibfield  {title} {\bibinfo {title} {Adiabatic quantum dynamics of the
  lipkin-meshkov-glick model},\ }\href
  {https://doi.org/10.1103/PhysRevB.78.104426} {\bibfield  {journal} {\bibinfo
  {journal} {Phys. Rev. B}\ }\textbf {\bibinfo {volume} {78}},\ \bibinfo
  {pages} {104426} (\bibinfo {year} {2008})}\BibitemShut {NoStop}%
\bibitem [{\citenamefont {Acevedo}\ \emph {et~al.}(2014)\citenamefont
  {Acevedo}, \citenamefont {Quiroga}, \citenamefont {Rodr\'{\i}guez},\ and\
  \citenamefont {Johnson}}]{Acevedo2014}%
  \BibitemOpen
  \bibfield  {author} {\bibinfo {author} {\bibfnamefont {O.~L.}\ \bibnamefont
  {Acevedo}}, \bibinfo {author} {\bibfnamefont {L.}~\bibnamefont {Quiroga}},
  \bibinfo {author} {\bibfnamefont {F.~J.}\ \bibnamefont {Rodr\'{\i}guez}},\
  and\ \bibinfo {author} {\bibfnamefont {N.~F.}\ \bibnamefont {Johnson}},\
  }\bibfield  {title} {\bibinfo {title} {New dynamical scaling universality for
  quantum networks across adiabatic quantum phase transitions},\ }\href
  {https://doi.org/10.1103/PhysRevLett.112.030403} {\bibfield  {journal}
  {\bibinfo  {journal} {Phys. Rev. Lett.}\ }\textbf {\bibinfo {volume} {112}},\
  \bibinfo {pages} {030403} (\bibinfo {year} {2014})}\BibitemShut {NoStop}%
\bibitem [{\citenamefont {Defenu}\ \emph {et~al.}(2018)\citenamefont {Defenu},
  \citenamefont {Enss}, \citenamefont {Kastner},\ and\ \citenamefont
  {Morigi}}]{Defenu2018}%
  \BibitemOpen
  \bibfield  {author} {\bibinfo {author} {\bibfnamefont {N.}~\bibnamefont
  {Defenu}}, \bibinfo {author} {\bibfnamefont {T.}~\bibnamefont {Enss}},
  \bibinfo {author} {\bibfnamefont {M.}~\bibnamefont {Kastner}},\ and\ \bibinfo
  {author} {\bibfnamefont {G.}~\bibnamefont {Morigi}},\ }\bibfield  {title}
  {\bibinfo {title} {Dynamical critical scaling of long-range interacting
  quantum magnets},\ }\href {https://doi.org/10.1103/PhysRevLett.121.240403}
  {\bibfield  {journal} {\bibinfo  {journal} {Phys. Rev. Lett.}\ }\textbf
  {\bibinfo {volume} {121}},\ \bibinfo {pages} {240403} (\bibinfo {year}
  {2018})}\BibitemShut {NoStop}%
\bibitem [{\citenamefont {Dutta}\ and\ \citenamefont
  {Dutta}(2017)}]{Anirban_dutta}%
  \BibitemOpen
  \bibfield  {author} {\bibinfo {author} {\bibfnamefont {A.}~\bibnamefont
  {Dutta}}\ and\ \bibinfo {author} {\bibfnamefont {A.}~\bibnamefont {Dutta}},\
  }\bibfield  {title} {\bibinfo {title} {Probing the role of long-range
  interactions in the dynamics of a long-range kitaev chain},\ }\href
  {https://doi.org/10.1103/PhysRevB.96.125113} {\bibfield  {journal} {\bibinfo
  {journal} {Phys. Rev. B}\ }\textbf {\bibinfo {volume} {96}},\ \bibinfo
  {pages} {125113} (\bibinfo {year} {2017})}\BibitemShut {NoStop}%
\bibitem [{\citenamefont {Defenu}\ \emph
  {et~al.}(2019{\natexlab{a}})\citenamefont {Defenu}, \citenamefont {Morigi},
  \citenamefont {Dell'Anna},\ and\ \citenamefont {Enss}}]{N_defenu_TS}%
  \BibitemOpen
  \bibfield  {author} {\bibinfo {author} {\bibfnamefont {N.}~\bibnamefont
  {Defenu}}, \bibinfo {author} {\bibfnamefont {G.}~\bibnamefont {Morigi}},
  \bibinfo {author} {\bibfnamefont {L.}~\bibnamefont {Dell'Anna}},\ and\
  \bibinfo {author} {\bibfnamefont {T.}~\bibnamefont {Enss}},\ }\bibfield
  {title} {\bibinfo {title} {Universal dynamical scaling of long-range
  topological superconductors},\ }\href
  {https://doi.org/10.1103/PhysRevB.100.184306} {\bibfield  {journal} {\bibinfo
   {journal} {Phys. Rev. B}\ }\textbf {\bibinfo {volume} {100}},\ \bibinfo
  {pages} {184306} (\bibinfo {year} {2019}{\natexlab{a}})}\BibitemShut
  {NoStop}%
\bibitem [{\citenamefont {del Campo}(2018)}]{delCampo2018}%
  \BibitemOpen
  \bibfield  {author} {\bibinfo {author} {\bibfnamefont {A.}~\bibnamefont {del
  Campo}},\ }\bibfield  {title} {\bibinfo {title} {Universal statistics of
  topological defects formed in a quantum phase transition},\ }\href
  {https://doi.org/10.1103/PhysRevLett.121.200601} {\bibfield  {journal}
  {\bibinfo  {journal} {Phys. Rev. Lett.}\ }\textbf {\bibinfo {volume} {121}},\
  \bibinfo {pages} {200601} (\bibinfo {year} {2018})}\BibitemShut {NoStop}%
\bibitem [{\citenamefont {Białończyk}\ \emph {et~al.}(2021)\citenamefont
  {Białończyk}, \citenamefont {Gómez-Ruiz},\ and\ \citenamefont {del
  Campo}}]{delcampo2021}%
  \BibitemOpen
  \bibfield  {author} {\bibinfo {author} {\bibfnamefont {M.}~\bibnamefont
  {Białończyk}}, \bibinfo {author} {\bibfnamefont {F.~J.}\ \bibnamefont
  {Gómez-Ruiz}},\ and\ \bibinfo {author} {\bibfnamefont {A.}~\bibnamefont {del
  Campo}},\ }\bibfield  {title} {\bibinfo {title} {{Exact thermal properties of
  free-fermionic spin chains}},\ }\href
  {https://doi.org/10.21468/SciPostPhys.11.1.013} {\bibfield  {journal}
  {\bibinfo  {journal} {SciPost Phys.}\ }\textbf {\bibinfo {volume} {11}},\
  \bibinfo {pages} {013} (\bibinfo {year} {2021})}\BibitemShut {NoStop}%
\bibitem [{\citenamefont {King}\ \emph {et~al.}(2022)\citenamefont {King},
  \citenamefont {Suzuki}, \citenamefont {Raymond}, \citenamefont {Zucca},
  \citenamefont {Lanting}, \citenamefont {Altomare}, \citenamefont {Berkley},
  \citenamefont {Ejtemaee}, \citenamefont {Hoskinson}, \citenamefont {Huang},
  \citenamefont {Ladizinsky}, \citenamefont {MacDonald}, \citenamefont
  {Marsden}, \citenamefont {Oh}, \citenamefont {Poulin-Lamarre}, \citenamefont
  {Reis}, \citenamefont {Rich}, \citenamefont {Sato}, \citenamefont
  {Whittaker}, \citenamefont {Yao}, \citenamefont {Harris}, \citenamefont
  {Lidar}, \citenamefont {Nishimori},\ and\ \citenamefont {Amin}}]{King2022}%
  \BibitemOpen
  \bibfield  {author} {\bibinfo {author} {\bibfnamefont {A.~D.}\ \bibnamefont
  {King}}, \bibinfo {author} {\bibfnamefont {S.}~\bibnamefont {Suzuki}},
  \bibinfo {author} {\bibfnamefont {J.}~\bibnamefont {Raymond}}, \bibinfo
  {author} {\bibfnamefont {A.}~\bibnamefont {Zucca}}, \bibinfo {author}
  {\bibfnamefont {T.}~\bibnamefont {Lanting}}, \bibinfo {author} {\bibfnamefont
  {F.}~\bibnamefont {Altomare}}, \bibinfo {author} {\bibfnamefont {A.~J.}\
  \bibnamefont {Berkley}}, \bibinfo {author} {\bibfnamefont {S.}~\bibnamefont
  {Ejtemaee}}, \bibinfo {author} {\bibfnamefont {E.}~\bibnamefont {Hoskinson}},
  \bibinfo {author} {\bibfnamefont {S.}~\bibnamefont {Huang}}, \bibinfo
  {author} {\bibfnamefont {E.}~\bibnamefont {Ladizinsky}}, \bibinfo {author}
  {\bibfnamefont {A.~J.~R.}\ \bibnamefont {MacDonald}}, \bibinfo {author}
  {\bibfnamefont {G.}~\bibnamefont {Marsden}}, \bibinfo {author} {\bibfnamefont
  {T.}~\bibnamefont {Oh}}, \bibinfo {author} {\bibfnamefont {G.}~\bibnamefont
  {Poulin-Lamarre}}, \bibinfo {author} {\bibfnamefont {M.}~\bibnamefont
  {Reis}}, \bibinfo {author} {\bibfnamefont {C.}~\bibnamefont {Rich}}, \bibinfo
  {author} {\bibfnamefont {Y.}~\bibnamefont {Sato}}, \bibinfo {author}
  {\bibfnamefont {J.~D.}\ \bibnamefont {Whittaker}}, \bibinfo {author}
  {\bibfnamefont {J.}~\bibnamefont {Yao}}, \bibinfo {author} {\bibfnamefont
  {R.}~\bibnamefont {Harris}}, \bibinfo {author} {\bibfnamefont {D.~A.}\
  \bibnamefont {Lidar}}, \bibinfo {author} {\bibfnamefont {H.}~\bibnamefont
  {Nishimori}},\ and\ \bibinfo {author} {\bibfnamefont {M.~H.}\ \bibnamefont
  {Amin}},\ }\bibfield  {title} {\bibinfo {title} {Coherent quantum annealing
  in a programmable 2,000-qubit ising chain},\ }\href
  {https://doi.org/10.1038/s41567-022-01741-6} {\bibfield  {journal} {\bibinfo
  {journal} {Nature Physics}\ }\textbf {\bibinfo {volume} {18}},\ \bibinfo
  {pages} {1324} (\bibinfo {year} {2022})}\BibitemShut {NoStop}%
\bibitem [{\citenamefont {Singh}\ \emph {et~al.}(2023)\citenamefont {Singh},
  \citenamefont {Dhara},\ and\ \citenamefont {Gangadharaiah}}]{MSingh_2023}%
  \BibitemOpen
  \bibfield  {author} {\bibinfo {author} {\bibfnamefont {M.}~\bibnamefont
  {Singh}}, \bibinfo {author} {\bibfnamefont {S.}~\bibnamefont {Dhara}},\ and\
  \bibinfo {author} {\bibfnamefont {S.}~\bibnamefont {Gangadharaiah}},\
  }\bibfield  {title} {\bibinfo {title} {Driven one-dimensional noisy kitaev
  chain},\ }\href {https://doi.org/10.1103/PhysRevB.107.014303} {\bibfield
  {journal} {\bibinfo  {journal} {Phys. Rev. B}\ }\textbf {\bibinfo {volume}
  {107}},\ \bibinfo {pages} {014303} (\bibinfo {year} {2023})}\BibitemShut
  {NoStop}%
\bibitem [{\citenamefont {Gherardini}\ \emph {et~al.}(2024)\citenamefont
  {Gherardini}, \citenamefont {Buffoni},\ and\ \citenamefont
  {Defenu}}]{Defenu2024}%
  \BibitemOpen
  \bibfield  {author} {\bibinfo {author} {\bibfnamefont {S.}~\bibnamefont
  {Gherardini}}, \bibinfo {author} {\bibfnamefont {L.}~\bibnamefont
  {Buffoni}},\ and\ \bibinfo {author} {\bibfnamefont {N.}~\bibnamefont
  {Defenu}},\ }\bibfield  {title} {\bibinfo {title} {Universal defects
  statistics with strong long-range interactions},\ }\href
  {https://doi.org/10.1103/PhysRevLett.133.113401} {\bibfield  {journal}
  {\bibinfo  {journal} {Phys. Rev. Lett.}\ }\textbf {\bibinfo {volume} {133}},\
  \bibinfo {pages} {113401} (\bibinfo {year} {2024})}\BibitemShut {NoStop}%
\bibitem [{\citenamefont {Fey}\ and\ \citenamefont
  {Schmidt}(2016)}]{SchmidtKaiPhillip2016}%
  \BibitemOpen
  \bibfield  {author} {\bibinfo {author} {\bibfnamefont {S.}~\bibnamefont
  {Fey}}\ and\ \bibinfo {author} {\bibfnamefont {K.~P.}\ \bibnamefont
  {Schmidt}},\ }\bibfield  {title} {\bibinfo {title} {Critical behavior of
  quantum magnets with long-range interactions in the thermodynamic limit},\
  }\href {https://doi.org/10.1103/PhysRevB.94.075156} {\bibfield  {journal}
  {\bibinfo  {journal} {Phys. Rev. B}\ }\textbf {\bibinfo {volume} {94}},\
  \bibinfo {pages} {075156} (\bibinfo {year} {2016})}\BibitemShut {NoStop}%
\bibitem [{\citenamefont {Sadhukhan}\ \emph {et~al.}(2020)\citenamefont
  {Sadhukhan}, \citenamefont {Sinha}, \citenamefont {Francuz}, \citenamefont
  {Stefaniak}, \citenamefont {Rams}, \citenamefont {Dziarmaga},\ and\
  \citenamefont {Zurek}}]{DSadhukhan_2020}%
  \BibitemOpen
  \bibfield  {author} {\bibinfo {author} {\bibfnamefont {D.}~\bibnamefont
  {Sadhukhan}}, \bibinfo {author} {\bibfnamefont {A.}~\bibnamefont {Sinha}},
  \bibinfo {author} {\bibfnamefont {A.}~\bibnamefont {Francuz}}, \bibinfo
  {author} {\bibfnamefont {J.}~\bibnamefont {Stefaniak}}, \bibinfo {author}
  {\bibfnamefont {M.~M.}\ \bibnamefont {Rams}}, \bibinfo {author}
  {\bibfnamefont {J.}~\bibnamefont {Dziarmaga}},\ and\ \bibinfo {author}
  {\bibfnamefont {W.~H.}\ \bibnamefont {Zurek}},\ }\bibfield  {title} {\bibinfo
  {title} {Sonic horizons and causality in phase transition dynamics},\ }\href
  {https://doi.org/10.1103/PhysRevB.101.144429} {\bibfield  {journal} {\bibinfo
   {journal} {Phys. Rev. B}\ }\textbf {\bibinfo {volume} {101}},\ \bibinfo
  {pages} {144429} (\bibinfo {year} {2020})}\BibitemShut {NoStop}%
\bibitem [{\citenamefont {Lakkaraju}\ \emph {et~al.}(2022)\citenamefont
  {Lakkaraju}, \citenamefont {Ghosh}, \citenamefont {Sadhukhan},\ and\
  \citenamefont {Sen(De)}}]{Aditi_sen_LR_model}%
  \BibitemOpen
  \bibfield  {author} {\bibinfo {author} {\bibfnamefont {L.~G.~C.}\
  \bibnamefont {Lakkaraju}}, \bibinfo {author} {\bibfnamefont {S.}~\bibnamefont
  {Ghosh}}, \bibinfo {author} {\bibfnamefont {D.}~\bibnamefont {Sadhukhan}},\
  and\ \bibinfo {author} {\bibfnamefont {A.}~\bibnamefont {Sen(De)}},\
  }\bibfield  {title} {\bibinfo {title} {Mimicking quantum correlation of a
  long-range hamiltonian by finite-range interactions},\ }\href
  {https://doi.org/10.1103/PhysRevA.106.052425} {\bibfield  {journal} {\bibinfo
   {journal} {Phys. Rev. A}\ }\textbf {\bibinfo {volume} {106}},\ \bibinfo
  {pages} {052425} (\bibinfo {year} {2022})}\BibitemShut {NoStop}%
\bibitem [{\citenamefont {Huang}\ \emph {et~al.}(2024)\citenamefont {Huang},
  \citenamefont {Zou},\ and\ \citenamefont {Ding}}]{ChengxiangDing_2024}%
  \BibitemOpen
  \bibfield  {author} {\bibinfo {author} {\bibfnamefont {Y.-H.}\ \bibnamefont
  {Huang}}, \bibinfo {author} {\bibfnamefont {Y.-T.}\ \bibnamefont {Zou}},\
  and\ \bibinfo {author} {\bibfnamefont {C.}~\bibnamefont {Ding}},\ }\bibfield
  {title} {\bibinfo {title} {Dynamical relaxation of a long-range kitaev
  chain},\ }\href {https://doi.org/10.1103/PhysRevB.109.094309} {\bibfield
  {journal} {\bibinfo  {journal} {Phys. Rev. B}\ }\textbf {\bibinfo {volume}
  {109}},\ \bibinfo {pages} {094309} (\bibinfo {year} {2024})}\BibitemShut
  {NoStop}%
\bibitem [{\citenamefont {Defenu}\ \emph
  {et~al.}(2019{\natexlab{b}})\citenamefont {Defenu}, \citenamefont {Enss},\
  and\ \citenamefont {Halimeh}}]{N_defenu_LR_2019}%
  \BibitemOpen
  \bibfield  {author} {\bibinfo {author} {\bibfnamefont {N.}~\bibnamefont
  {Defenu}}, \bibinfo {author} {\bibfnamefont {T.}~\bibnamefont {Enss}},\ and\
  \bibinfo {author} {\bibfnamefont {J.~C.}\ \bibnamefont {Halimeh}},\
  }\bibfield  {title} {\bibinfo {title} {Dynamical criticality and domain-wall
  coupling in long-range hamiltonians},\ }\href
  {https://doi.org/10.1103/PhysRevB.100.014434} {\bibfield  {journal} {\bibinfo
   {journal} {Phys. Rev. B}\ }\textbf {\bibinfo {volume} {100}},\ \bibinfo
  {pages} {014434} (\bibinfo {year} {2019}{\natexlab{b}})}\BibitemShut
  {NoStop}%
\bibitem [{\citenamefont {Sinha}\ \emph {et~al.}(2020)\citenamefont {Sinha},
  \citenamefont {Sadhukhan}, \citenamefont {Rams},\ and\ \citenamefont
  {Dziarmaga}}]{dziarmaga_2020}%
  \BibitemOpen
  \bibfield  {author} {\bibinfo {author} {\bibfnamefont {A.}~\bibnamefont
  {Sinha}}, \bibinfo {author} {\bibfnamefont {D.}~\bibnamefont {Sadhukhan}},
  \bibinfo {author} {\bibfnamefont {M.~M.}\ \bibnamefont {Rams}},\ and\
  \bibinfo {author} {\bibfnamefont {J.}~\bibnamefont {Dziarmaga}},\ }\bibfield
  {title} {\bibinfo {title} {Inhomogeneity induced shortcut to adiabaticity in
  ising chains with long-range interactions},\ }\href
  {https://doi.org/10.1103/PhysRevB.102.214203} {\bibfield  {journal} {\bibinfo
   {journal} {Phys. Rev. B}\ }\textbf {\bibinfo {volume} {102}},\ \bibinfo
  {pages} {214203} (\bibinfo {year} {2020})}\BibitemShut {NoStop}%
\bibitem [{\citenamefont {Cherng}\ and\ \citenamefont
  {Levitov}(2006)}]{Levitov_2006}%
  \BibitemOpen
  \bibfield  {author} {\bibinfo {author} {\bibfnamefont {R.~W.}\ \bibnamefont
  {Cherng}}\ and\ \bibinfo {author} {\bibfnamefont {L.~S.}\ \bibnamefont
  {Levitov}},\ }\bibfield  {title} {\bibinfo {title} {Entropy and correlation
  functions of a driven quantum spin chain},\ }\href
  {https://doi.org/10.1103/PhysRevA.73.043614} {\bibfield  {journal} {\bibinfo
  {journal} {Phys. Rev. A}\ }\textbf {\bibinfo {volume} {73}},\ \bibinfo
  {pages} {043614} (\bibinfo {year} {2006})}\BibitemShut {NoStop}%
\bibitem [{\citenamefont {Singh}\ and\ \citenamefont
  {Gangadharaiah}(2021)}]{MSingh_2021}%
  \BibitemOpen
  \bibfield  {author} {\bibinfo {author} {\bibfnamefont {M.}~\bibnamefont
  {Singh}}\ and\ \bibinfo {author} {\bibfnamefont {S.}~\bibnamefont
  {Gangadharaiah}},\ }\bibfield  {title} {\bibinfo {title} {Driven quantum spin
  chain in the presence of noise: Anti-kibble-zurek behavior},\ }\href
  {https://doi.org/10.1103/PhysRevB.104.064313} {\bibfield  {journal} {\bibinfo
   {journal} {Phys. Rev. B}\ }\textbf {\bibinfo {volume} {104}},\ \bibinfo
  {pages} {064313} (\bibinfo {year} {2021})}\BibitemShut {NoStop}%
\bibitem [{\citenamefont {Cincio}\ \emph {et~al.}(2007)\citenamefont {Cincio},
  \citenamefont {Dziarmaga}, \citenamefont {Rams},\ and\ \citenamefont
  {Zurek}}]{Zurek2007}%
  \BibitemOpen
  \bibfield  {author} {\bibinfo {author} {\bibfnamefont {L.}~\bibnamefont
  {Cincio}}, \bibinfo {author} {\bibfnamefont {J.}~\bibnamefont {Dziarmaga}},
  \bibinfo {author} {\bibfnamefont {M.~M.}\ \bibnamefont {Rams}},\ and\
  \bibinfo {author} {\bibfnamefont {W.~H.}\ \bibnamefont {Zurek}},\ }\bibfield
  {title} {\bibinfo {title} {Entropy of entanglement and correlations induced
  by a quench: Dynamics of a quantum phase transition in the quantum ising
  model},\ }\href {https://doi.org/10.1103/PhysRevA.75.052321} {\bibfield
  {journal} {\bibinfo  {journal} {Phys. Rev. A}\ }\textbf {\bibinfo {volume}
  {75}},\ \bibinfo {pages} {052321} (\bibinfo {year} {2007})}\BibitemShut
  {NoStop}%
\bibitem [{\citenamefont {Roychowdhury}\ \emph {et~al.}(2021)\citenamefont
  {Roychowdhury}, \citenamefont {Moessner},\ and\ \citenamefont
  {Das}}]{KRC_2021}%
  \BibitemOpen
  \bibfield  {author} {\bibinfo {author} {\bibfnamefont {K.}~\bibnamefont
  {Roychowdhury}}, \bibinfo {author} {\bibfnamefont {R.}~\bibnamefont
  {Moessner}},\ and\ \bibinfo {author} {\bibfnamefont {A.}~\bibnamefont
  {Das}},\ }\bibfield  {title} {\bibinfo {title} {Dynamics and correlations at
  a quantum phase transition beyond kibble-zurek},\ }\href
  {https://doi.org/10.1103/PhysRevB.104.014406} {\bibfield  {journal} {\bibinfo
   {journal} {Phys. Rev. B}\ }\textbf {\bibinfo {volume} {104}},\ \bibinfo
  {pages} {014406} (\bibinfo {year} {2021})}\BibitemShut {NoStop}%
\bibitem [{\citenamefont {Dziarmaga}\ and\ \citenamefont
  {Rams}(2022)}]{MarekRams_2022}%
  \BibitemOpen
  \bibfield  {author} {\bibinfo {author} {\bibfnamefont {J.}~\bibnamefont
  {Dziarmaga}}\ and\ \bibinfo {author} {\bibfnamefont {M.~M.}\ \bibnamefont
  {Rams}},\ }\bibfield  {title} {\bibinfo {title} {Kink correlations,
  domain-size distribution, and emptiness formation probability after a
  kibble-zurek quench in the quantum ising chain},\ }\href
  {https://doi.org/10.1103/PhysRevB.106.014309} {\bibfield  {journal} {\bibinfo
   {journal} {Phys. Rev. B}\ }\textbf {\bibinfo {volume} {106}},\ \bibinfo
  {pages} {014309} (\bibinfo {year} {2022})}\BibitemShut {NoStop}%
\bibitem [{\citenamefont {Lee}\ \emph {et~al.}(2016)\citenamefont {Lee},
  \citenamefont {Joglekar},\ and\ \citenamefont {Richerme}}]{TonyLee2016}%
  \BibitemOpen
  \bibfield  {author} {\bibinfo {author} {\bibfnamefont {T.~E.}\ \bibnamefont
  {Lee}}, \bibinfo {author} {\bibfnamefont {Y.~N.}\ \bibnamefont {Joglekar}},\
  and\ \bibinfo {author} {\bibfnamefont {P.}~\bibnamefont {Richerme}},\
  }\bibfield  {title} {\bibinfo {title} {String order via floquet interactions
  in atomic systems},\ }\href {https://doi.org/10.1103/PhysRevA.94.023610}
  {\bibfield  {journal} {\bibinfo  {journal} {Phys. Rev. A}\ }\textbf {\bibinfo
  {volume} {94}},\ \bibinfo {pages} {023610} (\bibinfo {year}
  {2016})}\BibitemShut {NoStop}%
\bibitem [{\citenamefont {Paul}\ \emph {et~al.}(2024)\citenamefont {Paul},
  \citenamefont {Titum},\ and\ \citenamefont {Maghrebi}}]{Maghrebi2024}%
  \BibitemOpen
  \bibfield  {author} {\bibinfo {author} {\bibfnamefont {S.}~\bibnamefont
  {Paul}}, \bibinfo {author} {\bibfnamefont {P.}~\bibnamefont {Titum}},\ and\
  \bibinfo {author} {\bibfnamefont {M.}~\bibnamefont {Maghrebi}},\ }\bibfield
  {title} {\bibinfo {title} {Hidden quantum criticality and entanglement in
  quench dynamics},\ }\href {https://doi.org/10.1103/PhysRevResearch.6.L032003}
  {\bibfield  {journal} {\bibinfo  {journal} {Phys. Rev. Res.}\ }\textbf
  {\bibinfo {volume} {6}},\ \bibinfo {pages} {L032003} (\bibinfo {year}
  {2024})}\BibitemShut {NoStop}%
\bibitem [{\citenamefont {Teretenkov}\ and\ \citenamefont
  {Lychkovskiy}(2024)}]{Teretenkov2024}%
  \BibitemOpen
  \bibfield  {author} {\bibinfo {author} {\bibfnamefont {A.}~\bibnamefont
  {Teretenkov}}\ and\ \bibinfo {author} {\bibfnamefont {O.}~\bibnamefont
  {Lychkovskiy}},\ }\bibfield  {title} {\bibinfo {title} {Exact dynamics of
  quantum dissipative xx models: Wannier-stark localization in the fragmented
  operator space},\ }\href {https://doi.org/10.1103/PhysRevB.109.L140302}
  {\bibfield  {journal} {\bibinfo  {journal} {Phys. Rev. B}\ }\textbf {\bibinfo
  {volume} {109}},\ \bibinfo {pages} {L140302} (\bibinfo {year}
  {2024})}\BibitemShut {NoStop}%
\bibitem [{\citenamefont {Mi}\ \emph {et~al.}(2024)\citenamefont {Mi},
  \citenamefont {Michailidis}, \citenamefont {Shabani}, \citenamefont {Miao},
  \citenamefont {Klimov},\ and\ \citenamefont {et~al.}}]{Mi2024}%
  \BibitemOpen
  \bibfield  {author} {\bibinfo {author} {\bibfnamefont {X.}~\bibnamefont
  {Mi}}, \bibinfo {author} {\bibfnamefont {A.~A.}\ \bibnamefont {Michailidis}},
  \bibinfo {author} {\bibfnamefont {S.}~\bibnamefont {Shabani}}, \bibinfo
  {author} {\bibfnamefont {K.~C.}\ \bibnamefont {Miao}}, \bibinfo {author}
  {\bibfnamefont {P.~V.}\ \bibnamefont {Klimov}},\ and\ \bibinfo {author}
  {\bibnamefont {et~al.}},\ }\bibfield  {title} {\bibinfo {title} {Stable
  quantum-correlated many-body states through engineered dissipation},\ }\href
  {https://doi.org/10.1126/science.adh9932} {\bibfield  {journal} {\bibinfo
  {journal} {Science}\ }\textbf {\bibinfo {volume} {383}},\ \bibinfo {pages}
  {1332} (\bibinfo {year} {2024})}\BibitemShut {NoStop}%
\bibitem [{\citenamefont {Shiroishi}\ \emph {et~al.}(2001)\citenamefont
  {Shiroishi}, \citenamefont {Takahashi},\ and\ \citenamefont
  {Nishiyama}}]{Nishiyama2001}%
  \BibitemOpen
  \bibfield  {author} {\bibinfo {author} {\bibfnamefont {M.}~\bibnamefont
  {Shiroishi}}, \bibinfo {author} {\bibfnamefont {M.}~\bibnamefont
  {Takahashi}},\ and\ \bibinfo {author} {\bibfnamefont {Y.}~\bibnamefont
  {Nishiyama}},\ }\bibfield  {title} {\bibinfo {title} {Emptiness formation
  probability for the one-dimensional isotropic xy model},\ }\href
  {https://doi.org/10.1143/JPSJ.70.3535} {\bibfield  {journal} {\bibinfo
  {journal} {Journal of the Physical Society of Japan}\ }\textbf {\bibinfo
  {volume} {70}},\ \bibinfo {pages} {3535} (\bibinfo {year}
  {2001})}\BibitemShut {NoStop}%
\bibitem [{\citenamefont {Franchini}\ and\ \citenamefont
  {Abanov}(2005)}]{Abanov_Franchini_2005}%
  \BibitemOpen
  \bibfield  {author} {\bibinfo {author} {\bibfnamefont {F.}~\bibnamefont
  {Franchini}}\ and\ \bibinfo {author} {\bibfnamefont {A.~G.}\ \bibnamefont
  {Abanov}},\ }\bibfield  {title} {\bibinfo {title} {Asymptotics of toeplitz
  determinants and the emptiness formation probability for the xy spin chain},\
  }\href {https://doi.org/10.1088/0305-4470/38/23/002} {\bibfield  {journal}
  {\bibinfo  {journal} {Journal of Physics A: Mathematical and General}\
  }\textbf {\bibinfo {volume} {38}},\ \bibinfo {pages} {5069} (\bibinfo {year}
  {2005})}\BibitemShut {NoStop}%
\bibitem [{\citenamefont {Ares}\ and\ \citenamefont {Viti}(2020)}]{Ares_2020}%
  \BibitemOpen
  \bibfield  {author} {\bibinfo {author} {\bibfnamefont {F.}~\bibnamefont
  {Ares}}\ and\ \bibinfo {author} {\bibfnamefont {J.}~\bibnamefont {Viti}},\
  }\bibfield  {title} {\bibinfo {title} {Emptiness formation probability and
  painlevé v equation in the xy spin chain},\ }\href
  {https://doi.org/10.1088/1742-5468/ab5d0b} {\bibfield  {journal} {\bibinfo
  {journal} {Journal of Statistical Mechanics: Theory and Experiment}\ }\textbf
  {\bibinfo {volume} {2020}},\ \bibinfo {pages} {013105} (\bibinfo {year}
  {2020})}\BibitemShut {NoStop}%
\bibitem [{\citenamefont {Kandala}\ \emph {et~al.}(2019)\citenamefont
  {Kandala}, \citenamefont {Temme}, \citenamefont {C{\'o}rcoles}, \citenamefont
  {Mezzacapo}, \citenamefont {Chow},\ and\ \citenamefont
  {Gambetta}}]{Kandala2019}%
  \BibitemOpen
  \bibfield  {author} {\bibinfo {author} {\bibfnamefont {A.}~\bibnamefont
  {Kandala}}, \bibinfo {author} {\bibfnamefont {K.}~\bibnamefont {Temme}},
  \bibinfo {author} {\bibfnamefont {A.~D.}\ \bibnamefont {C{\'o}rcoles}},
  \bibinfo {author} {\bibfnamefont {A.}~\bibnamefont {Mezzacapo}}, \bibinfo
  {author} {\bibfnamefont {J.~M.}\ \bibnamefont {Chow}},\ and\ \bibinfo
  {author} {\bibfnamefont {J.~M.}\ \bibnamefont {Gambetta}},\ }\bibfield
  {title} {\bibinfo {title} {Error mitigation extends the computational reach
  of a noisy quantum processor},\ }\href
  {https://doi.org/10.1038/s41586-019-1040-7} {\bibfield  {journal} {\bibinfo
  {journal} {Nature}\ }\textbf {\bibinfo {volume} {567}},\ \bibinfo {pages}
  {491} (\bibinfo {year} {2019})}\BibitemShut {NoStop}%
\bibitem [{\citenamefont {Li}\ \emph {et~al.}(2023)\citenamefont {Li},
  \citenamefont {Wu}, \citenamefont {Mei}, \citenamefont {Yao}, \citenamefont
  {Lian}, \citenamefont {Cai}, \citenamefont {Wang}, \citenamefont {Qi},
  \citenamefont {Yao}, \citenamefont {He}, \citenamefont {Zhou},\ and\
  \citenamefont {Duan}}]{BWLi2023}%
  \BibitemOpen
  \bibfield  {author} {\bibinfo {author} {\bibfnamefont {B.-W.}\ \bibnamefont
  {Li}}, \bibinfo {author} {\bibfnamefont {Y.-K.}\ \bibnamefont {Wu}}, \bibinfo
  {author} {\bibfnamefont {Q.-X.}\ \bibnamefont {Mei}}, \bibinfo {author}
  {\bibfnamefont {R.}~\bibnamefont {Yao}}, \bibinfo {author} {\bibfnamefont
  {W.-Q.}\ \bibnamefont {Lian}}, \bibinfo {author} {\bibfnamefont {M.-L.}\
  \bibnamefont {Cai}}, \bibinfo {author} {\bibfnamefont {Y.}~\bibnamefont
  {Wang}}, \bibinfo {author} {\bibfnamefont {B.-X.}\ \bibnamefont {Qi}},
  \bibinfo {author} {\bibfnamefont {L.}~\bibnamefont {Yao}}, \bibinfo {author}
  {\bibfnamefont {L.}~\bibnamefont {He}}, \bibinfo {author} {\bibfnamefont
  {Z.-C.}\ \bibnamefont {Zhou}},\ and\ \bibinfo {author} {\bibfnamefont
  {L.-M.}\ \bibnamefont {Duan}},\ }\bibfield  {title} {\bibinfo {title}
  {Probing critical behavior of long-range transverse-field ising model through
  quantum kibble-zurek mechanism},\ }\href
  {https://doi.org/10.1103/PRXQuantum.4.010302} {\bibfield  {journal} {\bibinfo
   {journal} {PRX Quantum}\ }\textbf {\bibinfo {volume} {4}},\ \bibinfo {pages}
  {010302} (\bibinfo {year} {2023})}\BibitemShut {NoStop}%
\bibitem [{\citenamefont {Azses}\ \emph {et~al.}(2023)\citenamefont {Azses},
  \citenamefont {Dupont}, \citenamefont {Evert}, \citenamefont {Reagor},\ and\
  \citenamefont {Dalla~Torre}}]{DanielAzses2023}%
  \BibitemOpen
  \bibfield  {author} {\bibinfo {author} {\bibfnamefont {D.}~\bibnamefont
  {Azses}}, \bibinfo {author} {\bibfnamefont {M.}~\bibnamefont {Dupont}},
  \bibinfo {author} {\bibfnamefont {B.}~\bibnamefont {Evert}}, \bibinfo
  {author} {\bibfnamefont {M.~J.}\ \bibnamefont {Reagor}},\ and\ \bibinfo
  {author} {\bibfnamefont {E.~G.}\ \bibnamefont {Dalla~Torre}},\ }\bibfield
  {title} {\bibinfo {title} {Navigating the noise-depth tradeoff in adiabatic
  quantum circuits},\ }\href {https://doi.org/10.1103/PhysRevB.107.125127}
  {\bibfield  {journal} {\bibinfo  {journal} {Phys. Rev. B}\ }\textbf {\bibinfo
  {volume} {107}},\ \bibinfo {pages} {125127} (\bibinfo {year}
  {2023})}\BibitemShut {NoStop}%
\bibitem [{\citenamefont {Perrin}\ \emph {et~al.}(2024)\citenamefont {Perrin},
  \citenamefont {Scoquart}, \citenamefont {Pavlov},\ and\ \citenamefont
  {Gnezdilov}}]{Gnezdilov2024}%
  \BibitemOpen
  \bibfield  {author} {\bibinfo {author} {\bibfnamefont {H.}~\bibnamefont
  {Perrin}}, \bibinfo {author} {\bibfnamefont {T.}~\bibnamefont {Scoquart}},
  \bibinfo {author} {\bibfnamefont {A.~I.}\ \bibnamefont {Pavlov}},\ and\
  \bibinfo {author} {\bibfnamefont {N.~V.}\ \bibnamefont {Gnezdilov}},\ }\href
  {https://arxiv.org/abs/2407.04770} {\bibinfo {title} {Dynamic thermalization
  on noisy quantum hardware}} (\bibinfo {year} {2024}),\ \Eprint
  {https://arxiv.org/abs/2407.04770} {arXiv:2407.04770 [quant-ph]} \BibitemShut
  {NoStop}%
\bibitem [{\citenamefont {Teplitskiy}\ \emph {et~al.}(2024)\citenamefont
  {Teplitskiy}, \citenamefont {Kiss}, \citenamefont {Grossi},\ and\
  \citenamefont {Mandarino}}]{AntonioMandarino2024}%
  \BibitemOpen
  \bibfield  {author} {\bibinfo {author} {\bibfnamefont {D.}~\bibnamefont
  {Teplitskiy}}, \bibinfo {author} {\bibfnamefont {O.}~\bibnamefont {Kiss}},
  \bibinfo {author} {\bibfnamefont {M.}~\bibnamefont {Grossi}},\ and\ \bibinfo
  {author} {\bibfnamefont {A.}~\bibnamefont {Mandarino}},\ }\href
  {https://arxiv.org/abs/2410.06250} {\bibinfo {title} {Statistics of
  topological defects across a phase transition in a superconducting quantum
  processor}} (\bibinfo {year} {2024}),\ \Eprint
  {https://arxiv.org/abs/2410.06250} {arXiv:2410.06250 [quant-ph]} \BibitemShut
  {NoStop}%
\bibitem [{\citenamefont {Miessen}\ \emph {et~al.}(2024)\citenamefont
  {Miessen}, \citenamefont {Egger}, \citenamefont {Tavernelli},\ and\
  \citenamefont {Mazzola}}]{Miessen2024}%
  \BibitemOpen
  \bibfield  {author} {\bibinfo {author} {\bibfnamefont {A.}~\bibnamefont
  {Miessen}}, \bibinfo {author} {\bibfnamefont {D.~J.}\ \bibnamefont {Egger}},
  \bibinfo {author} {\bibfnamefont {I.}~\bibnamefont {Tavernelli}},\ and\
  \bibinfo {author} {\bibfnamefont {G.}~\bibnamefont {Mazzola}},\ }\bibfield
  {title} {\bibinfo {title} {Benchmarking digital quantum simulations above
  hundreds of qubits using quantum critical dynamics},\ }\href
  {https://doi.org/10.1103/PRXQuantum.5.040320} {\bibfield  {journal} {\bibinfo
   {journal} {PRX Quantum}\ }\textbf {\bibinfo {volume} {5}},\ \bibinfo {pages}
  {040320} (\bibinfo {year} {2024})}\BibitemShut {NoStop}%
\end{thebibliography}%
\end{document}